\documentclass{article}
\usepackage{fullpage}

\usepackage{booktabs} 

\usepackage{amsmath}
\usepackage{amssymb}
\usepackage{amsthm}
\usepackage{amsfonts}
\usepackage{graphicx}
\usepackage{mathtools}
\usepackage{hyperref}
\usepackage{xspace}
\usepackage{aliascnt}

\usepackage{tikz}
\usetikzlibrary{arrows}
\usetikzlibrary{shapes}
\usetikzlibrary{backgrounds}

\usepackage{epsfig}

\newcommand{\Int}{\mathbb{Z}}
\newcommand{\IntMod}[1]{\Int_{#1}}

\renewcommand{\gets}{\mathrel{\mathop:}=}
\newcommand{\nameq}{\equiv}
\newcommand{\acteq}{\longrightarrow}

\renewcommand{\mod}{{}\textnormal{\texttt{\;mod\;}}{}}

\newcommand{\textiff}{\textit{iff}\xspace}
\newcommand{\quicksec}[1]{\noindent{\bf #1.}\;}
\newcommand{\mysf}{\sf \small \mbox{}}
\newcommand{\funcname}[1]{\textnormal{\mysf {#1}}}
\newcommand{\funcall}[2]{\funcname{#1}(#2)}

\newcommand{\RecEnumLang}{\ensuremath{\Sigma_1^0}}
\newcommand{\CoRecEnumLang}{\ensuremath{\Pi_1^0}}

\newcommand{\SemiDec}[1][]{\RecEnumLang\ifx\empty#1\empty\else-#1\fi\xspace}
\newcommand{\CoSemiDec}[1][]{\CoRecEnumLang\ifx\empty#1\empty\else-#1\fi\xspace}

\mathchardef\breakingcomma\mathcode`\,
{\catcode`,=\active
 \gdef,{\breakingcomma\discretionary{}{}{}}
}
\newcommand{\mathlist}[1]{\mathcode`\,=\string"8000 #1}

\hypersetup{
 pdfborderstyle={/S/U/W 0.5}
}

\usepackage{paralist}

\newenvironment{itemize*}%
{\begin{compactitem}}%
{\end{compactitem}}

\newenvironment{enumerate*}%
{\begin{compactenum}}%
{\end{compactenum}}

\usepackage{floatflt}

\usepackage{algorithm}
\usepackage{algorithmic}
\usepackage{subfig}

\newcommand{\protosym}{p}
\newcommand{\prefun}[1]{\funcall{Pre}{#1}}
\newcommand{\UniEnabled}[2]{(#1,#2)\in\prefun{\delta}}

\newcommand{\UniNextState}[2]{\delta(#1,#2)}

\newtheorem{theorem}{Theorem}[section]
\newcommand{\mynewtheorem}[2]{
 \newaliascnt{#1}{theorem}
 \newtheorem{#1}[#1]{#2}
 \aliascntresetthe{#1}
 \expandafter\def\csname #1autorefname\endcsname{#2}
}

\mynewtheorem{lemma}{Lemma}
\mynewtheorem{corollary}{Corollary}
\mynewtheorem{example}{Example}
\theoremstyle{definition}
\mynewtheorem{problem}{Problem}
\mynewtheorem{definition}{Definition}
\mynewtheorem{remark}{Remark}
\mynewtheorem{observation}{Observation}

\usepackage{float}
\floatstyle{boxed}
\restylefloat{figure}

\floatstyle{plain}
\restylefloat{figure}

\begin{document}

\title{Verification and Synthesis of   Symmetric Uni-Rings   for Leads-To Properties}

\author{Ali~Ebnenasir\\ Department of Computer Science\\Michigan Technological University\\Houghton MI 49931}
\date{May 2019}


\maketitle

\begin{abstract}
This paper investigates the verification and synthesis of parameterized protocols that satisfy leadsto properties $R \leadsto Q$ on symmetric unidirectional rings (a.k.a. uni-rings) of deterministic and constant-space processes under no fairness and interleaving semantics, where $R$ and $Q$ are global state predicates. First, we show that verifying $R \leadsto Q$ for parameterized protocols on symmetric uni-rings is undecidable, even for deterministic and constant-space processes, and conjunctive state predicates. Then, we show that surprisingly synthesizing  symmetric uni-ring protocols that satisfy $R \leadsto Q$ is actually decidable. 
We identify necessary and sufficient conditions for the decidability of synthesis based on which we devise a sound and complete polynomial-time algorithm that takes the predicates $R$ and $Q$, and automatically generates a parameterized protocol  that satisfies $R \leadsto Q$ for unbounded (but finite) ring sizes. 
 Moreover, we present some decidability results for cases where leadsto is required from multiple distinct $R$ predicates to different $Q$ predicates. To demonstrate the practicality of our synthesis method, we synthesize some parameterized protocols, including agreement and parity protocols. 
\end{abstract}
\newpage

\section{Introduction}
\label{sec:intro}

This paper investigates the verification and synthesis of  parameterized protocols that satisfy {\it leadsto} properties $R \leadsto Q$ on symmetric unidirectional rings (a.k.a. {\it uni-rings}) of deterministic and constant-space processes under no fairness and interleaving semantics, where $R$ and $Q$ represent global state predicates. The significance of this problem is two-fold. First,  ring  is a simple, but important topology for distributed systems whose underlying communication graph  includes cycles (which is the case in many practical domains). Second, the  leadsto property $R \leadsto Q$ is a  critical liveness requirement in numerous contexts where system executions should guarantee eventual response (i.e., reaching the set of states $Q$) to specific stimuli (i.e., being in the set of states $R$). In a {\it symmetric} ring, the code of each process is generated from the code of a {\it template/representative} process by a simple variable renaming.  Moreover, the number of processes in the ring is unbounded (but finite).  In this paper, we first extend  Suzuki's  undecidability results of verifying Linear Temporal Logic (LTL) properties of uni-rings  \cite{suzuki1988proving} to the special case of verifying leadsto properties for symmetric uni-rings of deterministic and constant-space processes, and show that the verification problem remains undecidable. We then present a surprising result that, despite the undecidability of verification, synthesizing uni-rings that satisfy leadsto properties is actually decidable. This is a counterintuitive result as it is believed \cite {pr90} that the synthesis of distributed systems is harder than their verification.



 
Most existing synthesis  methods for parameterized protocols are extensions/variants of bounded and parameterized synthesis with a focus on  Temporal Logic (TL) properties \cite{handb90} for general communication topologies and under fairness assumptions. For example,  Finkbeiner and Schewe  \cite{finkbeiner2013bounded} formulate  the synthesis of fixed-size  protocols as a set of  constraints, and use Satisfiability Modulo Theory (SMT) solvers \cite{de2011satisfiability} to find a  protocol that is accepted by a Universal Co-Buchi Tree (UCT) automaton generated from TL specifications. The search for a protocol is conducted up to a certain bound in the state space of processes and/or their automata-theoretic product; i.e., {\it bounded synthesis}.  Jacobs and Bloem \cite{jacobs2012parameterized} extend bounded synthesis to parameterized protocols by identifying {\it cutoff} bounds, where a solution exists for a protocol with {\it cutoff} number of processes {\it iff} (if and only if)  a solution exists for the parameterized protocol with unbounded number of processes (a.k.a. {\it parameterized synthesis}). Then, they apply the SMT-based bounded synthesis for an instance of the problem with at most {\it cutoff} processes. QBF-based bounded synthesis \cite{gascon2014synthesis} takes an incomplete design (a.k.a. {\it sketch}) of a protocol  and uses bounded synthesis towards generating fault-tolerant protocols  in a bounded space. While  methods for verification and synthesis of parameterized programs inspire our work, they (1) are mainly based on bounded/parameterized synthesis from temporal logic specifications; (2) make assumptions about synchrony, weak/strong fairness and complete knowledge of the network for each process; (3) the iterative nature of bounded and parameterized synthesis makes them computationally expensive and sensitive to their input (in part due to using SMT solvers), and (4) mostly focus on  safety and local liveness properties that are specified in terms of the neighborhood of a proper subset of processes (e.g., progress of each process).


This paper puts forward a paradigm shift where we make a step towards developing a {\it topology and property-specific approach}. Our long-term objective is to create an extensible library of synthesis algorithms and tools that can efficiently generate parameterized protocols for specific elementary topologies (e.g., ring, tree, mesh, etc.) and basic temporal properties (e.g., leadsto, until, safety, etc.). Another component of this new paradigm includes methods that compose parameterized protocols with elementary topologies while preserving correctness, which is beyond the scope of this work. The core novelty of the proposed approach of this paper lies in identifying local characterizations of global properties (e.g., reachability, livelocks) towards enabling synthesis in the local state space of the template process for global correctness. Such local characterizations would enable  more efficient automated reasoning methods, where the time/space complexity of synthesis will be in terms of the size of the local state space of template processes instead of semi-decision procedures that conduct  backward/forward reachability analysis \cite{conchon2012cubicle}.



In the case of symmetric uni-rings that satisfy leadsto properties $R \leadsto Q$, we first show that verifying $R \leadsto Q$ remains undecidable even for protocols with constant-space and deterministic processes, and for  conjunctive state predicates $R= \forall i \in \IntMod{N}: r(x_{i-1},x_i)$ and $Q= \forall  i \in \IntMod{N}: q(x_{i-1},x_i)$, where $N$ denotes the number of processes in the ring, $ \IntMod{N}$ represents values modulo $N$, and $x_i$ captures an abstraction of all writeable variables of the template process $P_i$. (Conjunctive predicates may seem restricted but they have important  applications for many systems \cite{varPhD93,jpdcGouda96}.) We then show some negative results that satisfying $R \leadsto Q$ by reaching  states in $Q$ where $x_{i-1} \neq x_i$ is impossible. Intuitively, this negative result is an outcome of the impossibility of recovery to an $L$-coloring  in symmetric uni-rings from any state (due to Bernard {\it et al.}'s \cite{ipdpsBernard09}). Subsequently, we show that synthesizing a parameterized protocol that satisfies $R \leadsto Q$  is decidable {\it iff} (if and only if)  there are some distinct values $v, \gamma \in ${\it Dom}$(x_i)$ for which $q(\gamma, \gamma)$ and  $q(v, \gamma)$ hold. 
 Our necessary and sufficient conditions are based on the intuition that  a parameterized protocol on a uni-ring that satisfies $R \leadsto Q$  exists {\it iff} there is a sequence of steps for every process towards satisfying $q(\gamma,\gamma)$ locally, starting from a  state that satisfies $R$. We then provide a sound and complete algorithm that takes the  state predicates $R$ and $Q$, and generates the parameterized actions of the template process, if a solution exists. The time complexity of the proposed algorithm is polynomial in the state space of the template process, which is often a small value for constant-space processes.  We also show that synthesis remains decidable for cases where leadsto is required from the conjunction/disjunction of a set of  predicates $R_1, \cdots, R_k$, where $k >1$, to the conjunction/disjunction of two  predicates $Q_1$ and $Q_2$.  To demonstrate the practicality of our algorithm, we present a few case studies including a protocol that ensures reaching agreement in  uni-rings when processes of the ring disagree on a value, and a parity protocol that guarantees a common parity amongst the processes. We conjecture that the implementation of our algorithm will provide a highly efficient synthesis tool as our previous work \cite{tseEbnenasir19} on the synthesis of fault-tolerant parameterized uni-rings confirms our belief.

\quicksec{Organization}
\autoref{sec:prelim} presents some basic concepts. \autoref{sec:undec} shows that verifying leadsto on uni-rings is undecidable.  \autoref{sec:dec} identifies necessary and sufficient conditions for decidability of synthesizing symmetric uni-rings that satisfy leadsto properties.  \autoref{sec:cases} presents a few case studies, including an agreement and a parity protocol. \autoref{sec:related} discusses related work, and \autoref{sec:concl} summarizes our contributions and outlines some future work.

\section{Preliminaries}
\label{sec:prelim}

This section presents the definition of parameterized protocols and  their representation as  action graphs. Wlog, we use the term {\it protocol} to refer to finite-state symmetric  uni-rings as we conduct our investigation in the context of network protocols. Such rings are parameterized in the number of processes in the ring. 
A protocol $\protosym$ for a computer network includes $N > 1$ processes (finite-state machines), where each process $P_i$ has a finite set of readable and writeable variables. Any {\it local state} of a process (a.k.a. {\it locality/neighborhood}) is determined by a unique valuation of its readable variables. We assume that any writeable variable is also readable.
The {\it global state} of a protocol is defined by a snapshot of the local states of all processes. The {\it state space} of a protocol, denoted by $\Sigma$, is the universal set of all global states. A {\it state predicate} is a subset of $\Sigma$. A process {\it acts} (i.e., {\it transitions}) when it atomically updates its state based on its locality. The locality of a process is defined by the network topology. 
For example, in a uni-ring  consisting of $N$ processes, each process $P_i$ (where $i\in\IntMod{N}$, i.e., $0 \leq i \leq N-1$) has a neighbor/predecessor $P_{i-1}$, where subtraction and addition are  in modulo $N$.
We assume that  processes act one at a time (i.e., interleaving semantics). Thus, each {\it global transition} corresponds to the action of a single process from some global state.
An {\it execution/computation} of a protocol is a sequence of states $s_0, s_1, \dots, s_k$ where there is a transition from $s_i$ to $s_{i+1}$ for every $i\in\IntMod{k}$.
We consider  {\it parameterized} protocols that consist of  families of symmetric processes. Each family is represented by a {\it template} process from which the code of all family members is instantiated by a simple variable renaming/re-indexing. 
For instance, a symmetric uni-ring includes just one family for which we use triples of the form $(a,b,c)$ to denote actions $(x_{i-1} = a\land x_i=b \acteq x_i\gets c;)$ of the template process $P_i$. An action has two components; a {\it guard}, which is a Boolean expression in terms of readable variables and a {\it statement} that atomically updates the state (i.e., writeable variables) of the process once the guard evaluates to {\it true}; i.e., the action is {\it enabled}.

\begin{definition}[Transition Function]
\label{def:TransitionFun}
Let $P_i$ be any process with a state variable $x_i$ in a uni-ring protocol $\protosym$.
We define its transition function $\delta: \Sigma \times \Sigma \to \Sigma$ as a partial function such that $\delta(a,b) = c$ if and only if $P_i$ has an action $(x_{i-1} = a\land x_i=b \acteq x_i\gets c;)$.
In other words, $\delta$ can be used to define all actions of $P_i$ in the form of a single parametric action:
\[ (\UniEnabled{x_{i-1}}{x_i}) \acteq x_i\gets \UniNextState{x_{i-1}}{x_i}; \]
where $\UniEnabled{x_{i-1}}{x_i}$ checks to see if the current $x_{i-1}$ and $x_i$ values are in the preimage of $\delta$.  For other topologies, the same  definition of transition function holds except that  the preimage of $\delta$ might be specified differently depending on the locality of each process.
\end{definition}


\begin{definition}[Action Graph]
\label{def:actGraph}
We depict the set of actions of the template process of a  symmetric uni-ring by a labeled directed multigraph $G=(V, A)$, called the {\it action graph}, where each vertex $v \in V$ represents a value in $\IntMod M$, where $M$ denotes the domain size of $x_i$ and each arc $(a, c) \in A$ with a label $b$ captures an action  $x_{i-1} = a \land x_i=b \acteq x_i\gets c$. 
\end{definition}

For example, consider the Sum-Not-2 protocol given in~\cite{icdcsFarahatE12}.
Each process $P_i$ has a variable $x_i\in\IntMod{3}$ and actions
$x_{i-1} = 0 \land x_i = 2 \acteq x_i\gets 1$,
$x_{i-1} = 1 \land x_i = 1 \acteq x_i\gets 2$,
and
$x_{i-1} = 2 \land x_i = 0 \acteq x_i\gets 1$. This protocol ensures that, from any global state, a state is reached  where the sum of each two consecutive $x$ values does not equal $2$. The set of such states is formally specified as the state predicate $\forall i \in\IntMod{N}: (x_{i-1} + x_i \ne 2)$. 
We can represent this protocol as a graph containing arcs $(0,2,1)$, $(1,1,2)$, and $(2,0,1)$ as shown in \autoref{fig:SumNotTwo}.

\begin{floatingfigure}{5cm}
\centering
\begin{tikzpicture}[framed,x=2cm,y=1.5cm]
\tikzstyle{snod}=[draw,circle]
\draw
(0, 0) node[snod] (0) {$0$}
++(1, 0) node[snod] (1) {$1$}
++(1, 0) node[snod] (2) {$2$};
\tikzstyle{xn}=[draw,arrows=-latex,font=\scriptsize]
\path[xn]
(0) edge                node [above] {$2$} (1)
(1) edge [bend left=15] node [above] {$1$} (2)
(2) edge [bend left=15] node [below] {$0$} (1)
;
\end{tikzpicture}
\caption{Graph representing Sum-Not-2 protocol.}
\label{fig:SumNotTwo}
\end{floatingfigure}
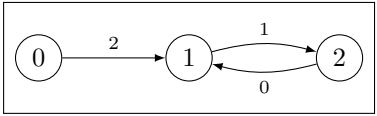

\ \\

For simplicity, we assume that  protocols consist of {\em self-disabling} processes. As such, an action $(a,b,c)$ cannot coexist with action $(a,c,d)$ for any $d$.
Moreover,  a {\it deterministic} process cannot have two actions enabled at the same time; i.e., an action $(a,b,c)$ cannot coexist with an action $(a,b,d)$ where $d\ne c$.

\begin{definition}[Leadsto Properties]
\label{def:leadsto}
The focus of this paper is on {\it leadsto} properties that are specified as $\Box (R \Rightarrow \Diamond Q)$ in Linear Temporal Logic (LTL), also denoted $R \leadsto Q$, where $\Box$ and $\Diamond$ respectively denote the universality and eventuality modalities, and $R = \forall i \in \IntMod{N}: r(x_{i-1},x_i)$ and  $Q = \forall  i \in \IntMod{N}: q(x_{i-1},x_i)$ represent conjunctive state predicates for a uni-ring of $N$ processes.  
A computation $\sigma = \langle s_0, s_1, \cdots \rangle$ of a protocol $p$ satisfies $\Box S$ {\em iff} the state predicate $S$ holds in every state $s_i \in \sigma$, for all $i \geq 0$.
A computation $\sigma = \langle s_0, s_1, \cdots \rangle$ of a protocol $p$ satisfies $\Diamond S$  {\em iff}  there is some $i \geq 0$ such that the state predicate $S$ holds in  the state $s_i \in \sigma$.
A computation $\sigma = \langle s_0, s_1, \cdots \rangle$ of a protocol $p$ satisfies $R \leadsto Q$  {\em iff}  $s_0 \in R$ implies that there is some $i \geq 0$ such that $s_i \in Q$.  A symmetric uni-ring protocol $p$ satisfies $R \leadsto Q$  {\em iff}  all the computations of $p$ satisfy $R \leadsto Q$, for unbounded (but finite) ring sizes.
\end{definition}

\begin{definition}[Fairness]
\label{def:fair}
A {\it strongly} fair scheduler ensures that any action that is infinitely often enabled will be executed infinitely often (due to Gouda \cite{wssGouda01}). A {\it weakly} fair scheduler guarantees that if an action is continuously enabled, then it will be executed infinitely often. 
\end{definition}

\quicksec{Livelock, deadlock, and closure}
A {\it global livelock} of a protocol $\protosym$ is an infinite computation $l = \langle s_0, s_1, \cdots, s_0 \rangle$, where $s_i$ is a global state, for all $i \geq 0$.  A {\it local livelock} of a process $P$ of protocol $\protosym$ is an infinite execution $l = \langle s_0, s_1, \cdots, s_0 \rangle$, where $s_i$ is a local state of $P$, for all $i \geq 0$. Unless stated otherwise, we use the terms `'livelock'' and `'global livelock'' interchangeably. For satisfying a leadsto property $R \leadsto Q$, a reachable livelock $l$ that includes at least one state in $Q$ is acceptable; otherwise, it is considered as a failure towards satisfying $R \leadsto Q$.
A {\it deadlock} of $\protosym$ is a state that has no outgoing transition; i.e., no process is enabled to act.
A state predicate $I$ is {\it closed} under $\protosym$ {\em iff} there is no transition $(s, s')$, where $s \in I$ but $s' \notin I$.

\begin{definition}[Self-Stabilization and Convergence]
\label{def:SS}
A protocol $\protosym$ is {\it self-stabilizing} \cite{dij} to a state predicate $I$ (under no fairness) {\em iff} from any  state in $\neg I$, {\em every} computation of $\protosym$ reaches a state in $I$ (i.e., {\it convergence}) and remains in $I$  (i.e., {\it closure}). That is, $\protosym$ is livelock-free and deadlock-free in $\neg I$, and $I$ is closed under $\protosym$. A protocol $\protosym$ is {\it silent-stabilizing} to $I$  {\em iff} $\protosym$ converges to $I$ and $p$ has no computation starting in $I$. Notice that, in LTL, convergence to $I$ is specified as $\Box \Diamond I$, which is logically equivalent to {\it true} $\leadsto I$.


\end{definition}

\begin{definition}[Weak Stabilization]
\label{def:SS}
 A {\it weakly} stabilizing protocol to $I$ ensures that from each  state in $\neg I$, there is {\em some} computation that reaches a state in $I$ (a.k.a., {\it weak convergence}) and remains in $I$. 
\end{definition}

Notice that, any self-stabilizing protocol is also weakly stabilizing but the reverse is not true.

\begin{definition}[Locality Graphs]
\label{def:lGraph}
Consider a state predicate $Q = \forall i: q(x_{i-1},x_i)$ for a uni-ring. The relation $q(x_{i-1},x_i)$ captures a set of local  states, representing an acceptable relation between each process $P_i$ and its predecessor. The relation $q(x_{i-1},x_i)$ must also be {\it locally correctable} in that, for any value of $x_{i-1}$, there is always a sequence of steps that $P_i$ can take to establish $q$  by updating $x_i$ only. 
To simplify reasoning, we  represent $q(x_{i-1},x_i)$ as a digraph $G=(V, A)$, called the {\it locality graph}, such that each vertex $v \in V$ represents a value in  $\IntMod M$, and an arc $(a,b)$ is in $A$ \textiff $q(a,b)$ is true.
\end{definition}

Figure \ref{fig:locGNot2} illustrates the locality graph of the Sum-Not-2 protocol introduced in this section for the state predicate $Q= \forall i \in\IntMod{N}: (x_{i-1} \oplus x_i \ne 2)$ where $M = 4$ and $\oplus$ represents addition modulo $4$. Each closed walk in the locality graph characterizes a class of global states in the state  predicate $Q$. For example, the  closed walk   $(0, 1, 3, 1, 0)$ captures the global states of ring sizes $4 \times k$ where the $x_i$ value of processes follow a repeated pattern of $0, 1, 3, 1$, and $k$ is a positive integer. We now represent one of our previous results \cite{icdcsFarahatE12} on the relation between closed walks in locality graphs and global states of parameterized uni-rings.

\newcommand{\TikzDecSyntInstance}{%
\tikzstyle{snod}=[draw,circle]
\tikzstyle{phantnod}=[inner sep=0pt]
\draw
( 0, 0) node[snod] (0) {$0$}
( 1, 0) node[snod] (1) {$1$}
( 0,-1) node[snod] (2) {$2$}
( 1,-1) node[snod] (3) {$3$}
;
\tikzstyle{lab}=[font=\scriptsize]
\tikzstyle{xn}=[draw,arrows=-latex,font=\scriptsize]}

\begin{figure}[h]
\centering
\makebox[.2\textwidth]{%
\begin{tikzpicture}[x=2cm,y=1.5cm]
\TikzDecSyntInstance
\draw[xn]
(0) edge[out=160,in=220,looseness=7]  (0)
(2) edge[out=160,in=220,looseness=7]  (2)
(0) edge[bend  left=15] (1)
(1) edge[bend  left=15] (0)
(0) edge[bend  left=15] (3)
(3) edge[bend  left=15] (0)
(2) edge[bend  left=15] (3)
(3) edge[bend  left=15] (2)
(1) edge[bend  left=15] (3)
(3) edge[bend  left=15] (1)
(1) edge[bend  left=15] (2)
(2) edge[bend  left=15] (1)
;
\end{tikzpicture}%
}%
\caption{Locality graph of the SumNotTwo protocol.}
\label{fig:locGNot2}
\end{figure}
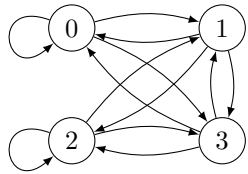

\begin{theorem}
\label{thm:cycleGState}
Any closed walk of length $L \geq 1$ in the locality graph of a conjunctive predicate $Q$ of a uni-ring characterizes global states in $Q$ of ring sizes of $L \times k$, where $k$ is a positive integer. (Proof in \cite{icdcsFarahatE12}) 
\end{theorem}

A corollary of Theorem \ref{thm:cycleGState} is as follows:

\begin{corollary}
\label{cor:acycl}
Any conjunctive state predicate $Z= \forall i : i \in \IntMod{N}: z(x_{i-1},x_i)$ whose locality graph $G_Z$ is acyclic specifies an empty set of states in a parameterized symmetric uni-ring.
\end{corollary}

\noindent Next, we represent some of our previous results regarding livelocks (from \cite{icdcsFarahatE12,sssKlink13}) that we shall use in this paper.

\quicksec{Propagations and Collisions}
When a process acts and enables its successor in a uni-ring, it propagates its ability to act.
The successor may enable its own successor by acting, and the pattern may continue indefinitely.
Such behaviors can be represented as sequences of actions that are propagated in a ring, called {\it propagations}. A propagation is a walk through the action graph.
For example, the Sum-Not-2 protocol has a propagation $\langle (0,2,1), (1,1,2), (2,0,1), (1,1,2) \rangle$ whose actions can be executed in order by processes $P_i$, $P_{i+1}$, $P_{i+2}$, and $P_{i+3}$ from a state
$(\mathlist{x_{i-1}, x_{i}, x_{i+1}, x_{i+2}, x_{i+3}}) = (\mathlist{0,2,1,0,1})$.
A propagation is {\em periodic} with period $n$ {\em iff} its $j$-th action and $(j+n)$-th action are the same for every index $j$.
A  propagation with period $n \geq 1$ corresponds to a {\it closed walk} of length $n$ in the action graph.
The Sum-Not-2 protocol has such a propagation of period $2$: $\langle (1,1,2), (2,0,1) \rangle$ (see Figure \ref{fig:SumNotTwo}). A {\it collision} occurs when two consecutive processes, say $P_i$ and $P_{i+1}$, have  enabled actions; e.g., $(a,b,c)$ and $(b,e,f)$, where $b \neq c$. In such a scenario, $x_{i-1} {=} a, x_i {=} b, x_{i+1} {=} e$. A collision occurs when $P_i$ executes and assigns $c$ to $x_i$. If that occurs, $P_i$ will be disabled (because  processes are self-disabling), and $P_{i+1}$ becomes disabled too because $x_i$ is no longer equal to $b$. As a result, two enabled processes become disabled by one action.

\quicksec{``Leads'' Relation}
Consider two actions $A_1$ and $A_2$ in a process $P_i$.
We say the action $A_1$ {\em leads} $A_2$  {\em iff} the value of the variable $x_i$ after executing $A_1$ is the same as the value required for $P_i$ to execute $A_2$.
Formally, this means an action $(a,b,c)$ leads $(d,e,f)$  {\em iff} $e = c$.
Similarly, a propagation leads another  {\em iff} for every index $j$, its $j$-th action leads the $j$-th action of the other propagation.
In the action graph, this corresponds to two walks  where the label of the destination node of the $j$-th arc in the first walk matches the arc label of the $j$-th arc in the second walk (for each index $j$). In  \cite{sssKlink13,livelock2019}, we prove the following theorem:

\begin{theorem}
\label{thm:LivelockIFF}
A uni-ring protocol of symmetric, deterministic and self-disabling processes has a livelock for some ring size  {\em iff}  there exist  $m$ propagations with some period $n $, where the $(i-1)$-th propagation leads the $i$-th propagation for each index $i$ modulo $m$, for $m > 1$ and $n \geq 1$; i.e., the propagations {\it successively} lead each other modulo $m$.
\end{theorem}

We have shown \cite{icdcsFarahatE12} that verifying deadlock-freedom in uni-rings is decidable. 
However, checking livelock-freedom is an undecidable problem for uni-ring protocols (with self-disabling and deterministic processes) \cite{sssKlink13,livelock2019}. 


\begin{theorem}
\label{undecLive}
Verifying livelock-freedom in a symmetric uni-ring protocol (with self-disabling and deterministic processes) is undecidable \cite{sssKlink13,livelock2019}.
\end{theorem}

 We have also shown that verifying livelock-freedom remains undecidable even for a special type of livelocks, where exactly one process is enabled in every state of the livelocked computation; i.e., {\it deterministic livelocks} \cite{sssKlink13}.

\begin{theorem}
\label{undecDetLive}
Verifying livelock-freedom in a symmetric uni-ring protocol (with self-disabling and deterministic processes) remains undecidable even for deterministic livelocks \cite{sssKlink13}.
\end{theorem}

Since in every state of a deterministic livelock there is exactly one enabled process, the choice of fairness policy has no impact on which process will be executed in each state because the scheduler has only one enabled process to select.

\begin{corollary}
\label{undecDetLiveFair}
Verifying livelock-freedom in a symmetric uni-ring protocol (with self-disabling and deterministic processes) remains undecidable regardless of the fairness assumption (i.e., scheduling policy) \cite{sssKlink13}.
\end{corollary}

The above results imply the undecidability of verifying self-stabilization for symmetric uni-rings.

\begin{theorem}
\label{undecSS}
Verifying silent-stabilization (and self-stabilization) for a symmetric uni-ring protocol (with self-disabling and deterministic processes) is undecidable \cite{sssKlink13}.
\end{theorem}

While verifying self-stabilization for uni-rings is undecidable, we have shown that synthesis of self-stabilizing uni-rings is surprisingly decidable.

\begin{theorem}
\label{decSS}
Synthesizing silent-stabilization for a parameterized uni-ring protocol (with self-disabling, deterministic and constant-space processes) is decidable \cite{fsenKlink17,tseEbnenasir19}. 
\end{theorem}

\begin{theorem}
\label{decSynSS}
Let $x_i$ denote  a variable  representing the local state space of each process $P_i$ in a symmetric uni-ring of $N$ processes (where $i \in \IntMod{N}$), and the domain size of $x_i$ be a fixed value regardless of $N$. Then,  there is a  protocol $p$  that  self-stabilizes to a state predicate $Q=\forall i \in \IntMod{N}: q(x_{i-1},x_i)$ for unbounded (but finite) ring size $N$ {\it iff} there is a vertex $\gamma$ in the locality graph of $Q$, where $\gamma$ has a self-loop (i.e.,  $ q(\gamma,\gamma)$ holds) that can be reached from another vertex.
\end{theorem}

\section{Undecidability of Verification}
\label{sec:undec}

This section presents some impossibility results for the verification and synthesis of symmetric uni-ring protocols that satisfy   {\it leads-to} properties. Throughout the rest of the paper, $R = \forall  i \in \IntMod{N}: r(x_{i-1},x_i)$  and $Q=\forall  i \in \IntMod{N}: q(x_{i-1},x_i)$ represent conjunctive state predicates.   First, we formulate the verification problem as follows:

 \begin{problem} [Verification of LeadsTo]
 \label{verifProb}
 Let $p$ be a symmetric parameterized protocol on a  uni-ring, and $R$ and $Q$ represent conjunctive state predicates. Does $p$ satisfy  $R \leadsto Q$ for unbounded (but finite) ring sizes?  
 \end{problem} 
 
 
 
 
 \begin{theorem}
 Problem \ref{verifProb} is undecidable for  uni-rings  of  self-disabling, constant-space and deterministic processes. 
  \end{theorem} 
  
   \begin{proof}
 For a symmetric protocol $p$ to satisfy $R \leadsto Q$ on a uni-ring, $p$ should ensure that starting in $R$, its computations will eventually reach some state in $Q$ under no fairness and interleaving semantics. This requires deadlock and livelock-freedom of computations of $p$ that start in $R$. Due to undecidability of verifying livelock-freedom on symmetric uni-rings (Theorem \ref{undecLive}),  Problem \ref{verifProb} is also undecidable.
   \end{proof} 
   

 \begin{problem} [Synthesis of LeadsTo]
 \label{synthProb}
 Let $R$ and $Q$ be conjunctive state predicates. Consider a  symmetric uni-ring of self-disabling, constant-space and deterministic processes. Does there exist a symmetric protocol $p$  on the ring that satisfies $R \leadsto Q$ for unbounded (but finite)  ring sizes?
 \end{problem} 
 
 To investigate Problem \ref{synthProb}, we first consider a special case of this problem where $R$={\it true}.  
 
  \begin{problem} [Synthesis of Convergence]
 \label{specSynthProb}
 Let $Q$ be a conjunctive state predicate. Consider a  symmetric uni-ring of self-disabling, constant-space and deterministic processes. Does there exist a symmetric protocol $p$  on the ring that satisfies $true \leadsto Q$ for unbounded (but finite) ring sizes?
 \end{problem}

Converging to a state where $Q= \forall i \in \IntMod{N}: q(x_{i-1},x_i)$ holds in a uni-ring can be achieved in two ways. First, there may be some value $c \in \IntMod{M}$ such that $q(c,c)$ holds. Such values represent themselves as self-loops in the locality graph of $q(x_{i-1},x_i)$. Second, $L$ values $c_0, c_1, \cdots, c_{L-1} \in \IntMod{M}$ satisfy $ q(x_{i-1},x_i)$ in a circular fashion, where $ q(c_0,c_1), q(c_1, c_2), \cdots, q(c_{L-2},c_{L-1}), q(c_{L-1},c_0)$ hold and  $L>1$. Such values form a cycle of length $L$ in the locality graph of $q(x_{i-1},x_i)$. A cycle like that represents a family of rings that include global states where $Q$ holds through  an ordered placement of the values that appear in the cycle. The sizes of such rings are multiples of $L$; i.e., ring sizes of $k \cdot L$, where $k$ is a positive integer (Theorem \ref{thm:cycleGState}). To design a parameterized symmetric protocol $p$ that satisfies $\Diamond Q$ from some initial states captured by a state predicate $R$, developers should design $p$ in such a way that it ensures convergence to one of the aforementioned scenarios (under no fairness and interleaving semantics). We first show that, there is no protocol that can ensure  convergence through the second scenario.

\begin{definition} [Ordered $L$-coloring]
Consider  a set of $L$ distinct colors and a permutation function $next(c)$ (i.e., a bijective function from $L$ to $L$) that takes a color $c \in L$ and returns a color $c' \in L$ where $c \neq c'$. An ordered $L$-coloring of a uni-ring  is an $L$-coloring where $\forall i \in \IntMod{N}: c_i = next(c_{i \ominus 1})$,  $N$ denotes the number of processes in the ring, and $\ominus$ represents subtraction modulo $N$.
\end{definition}

\begin{theorem} 
\label{Lcoloring}
No protocol exists that converges to an ordered $L$-coloring  on symmetric uni-rings for rings of $N>L$ processes. 
\end{theorem}

\begin{proof}  
Bernard {\it et al.}'s \cite{ipdpsBernard09} show that no converging $L$-coloring protocol exists on symmetric uni-rings for rings of $N>L$ processes. By contradiction, assume there is a converging  protocol $p$ that ensures an {\em ordered} $L$-coloring for some uni-ring  with size $N > L$. Such an ordered $L$-coloring is a legitimate $L$-coloring. Thus, $p$ converges to an $L$-coloring, which is a contradiction with  \cite{ipdpsBernard09}.
\end{proof}

\begin{corollary}
\label{Qcoloring-fair}
Theorem \ref{Lcoloring} holds under any fairness assumption, including strong fairness. In other words,  Theorem \ref{Lcoloring} holds for weak convergence too.
\end{corollary}

\begin{proof} 
By contradiction, suppose Theorem \ref{Lcoloring} is falsified assuming strong fairness. That is, there is a weakly converging  protocol that guarantees $L$-coloring on uni-rings of $N>L$ processes. This means that from any arbitrary state, there is at least a computation that reaches a valid $L$-coloring of the uni-ring for any  $N>L$. Wlog, consider the case where $N = L +1$; i.e., the number of colors is one unit less than the number of processes, and processes that have a similar color to their predecessor simply choose the next color available by incrementing their value. Bernard {\it et al.} \cite{ipdpsBernard09} show that the executions of any deterministic coloring protocol are isomorphic to a coloring protocol where all processes follow the rule of incrementing their value (i.e., choosing the next color in order). Now, consider an example state $\langle 0, 0, 1, 2, 3, \cdots, N-2 \rangle$ for the uni-ring of size $N$. In this state, only the second process is enabled to change its color because it is the only processes that has the same color as that of its predecessor's. The execution of the second process would get the protocol to the state $\langle 0, 1, 1, 2, 3, \cdots, N-2 \rangle$, where the third process is now enabled. Following the same pattern, the uni-ring reaches the state $\langle 0, 1, 2, 2, 3, \cdots, N-2 \rangle$, where the fourth process is the only process that is enabled. This sequence of states forms the livelock   $\langle 0, 0, 1, 2, 3, \cdots, N-2 \rangle$,  $\langle 0, 1, 1, 2, 3, \cdots, N-2 \rangle$, $\langle 0, 1, 2, 2, 3, \cdots, N-2 \rangle$, $\cdots$, $\langle 0, 1, 2, 3, 4, \cdots, N-2, N-2 \rangle$, $\langle  0, 1, 2, 3, \cdots, N-2, 0 \rangle$. Notice that, the last state is actually the same as the initial state of this execution of the symmetric uni-ring. Moreover, this livelock is deterministic. Thus, by Corollary \ref{undecDetLiveFair}, even strong fairness will not resolve this deterministic livelock, which prevents convergence to $L$-coloring. A similar argument holds for weak fairness.
\end{proof} 



\begin{theorem}
\label{Qcoloring}
Let $Q$ be a conjunctive state predicate $\forall  i \in \IntMod{N}: q(x_{i-1},x_i)$. No  symmetric uni-ring protocol exists that can converge to states where $Q$ holds through cyclic satisfaction of $ q(c_0,c_1), q(c_1, c_2), \cdots, q(c_{L-2},c_{L-1}), q(c_{L-1},c_0)$, for rings of $N>L$ processes and $L>1$. This result holds under any fairness assumption. 
\end{theorem}

\begin{proof}
By contradiction, let $p$ be a parameterized  protocol that converges to states in $Q$ (under any fairness assumption) where  $ q(c_0,c_1), q(c_1, c_2), \cdots, q(c_{L-2},c_{L-1}), q(c_{L-1},c_0)$ hold for some values $c_0, c_1, \cdots, c_{L-1}$, where $L>1$. Thus, the locality graph of $ q(x_{i-1},x_i)$ must include a cycle whose vertices are labeled with $c_0, c_1, \cdots, c_{L-1}$ in order. As a results, such states of $Q$ represent an ordered $L$-coloring. That is, $p$ converges to an ordered $L$-coloring protocol for $N > L$. This is a contradiction with Theorem \ref{Lcoloring} and Corollary \ref{Qcoloring-fair}.
\end{proof}


\begin{theorem}
\label{undecEventual}
Let  $R$ and $Q$ be non-empty conjunctive state predicates,  $R \cap Q = \emptyset$ and $R \neq true$. No symmetric protocol exists on uni-rings that can satisfy $(R \leadsto  Q)$ by reaching states where $Q$ holds through cyclic satisfaction of $ q(c_0,c_1), q(c_1, c_2), \cdots, q(c_{L-2},c_{L-1}), q(c_{L-1},c_0)$, for rings of $N>L$ processes.
\end{theorem}

\begin{proof}
By contradiction, let $p$ be such a protocol whose computations reach states in $Q$ from states in $R$ through cyclic satisfaction of $ q(c_0,c_1), q(c_1, c_2), \cdots, q(c_{L-2},c_{L-1}), q(c_{L-1},c_0)$.   Using the decidability of synthesizing self-stabilizing protocols (Theorem \ref{decSynSS}), we  design a silent stabilizing protocol $p'$ that converges to $R$ from any state, and once in $R$, $p'$ becomes disabled. However, the actions of $p$ and $p'$ may interfere by creating livelocks outside $R \vee Q$. Such livelocks include states where $p$ and $p'$  both have enabled processes that can take some action. Thus, such livelocks are not deterministic livelocks. To ensure recovery from such livelocks,  we assume strong fairness. As a result, we guarantee that $p'$ will eventually converge to $R$, and from $R$, protocol $p$ can guarantee that we reach states in $Q$  through cyclic satisfaction of $ q(c_0,c_1), q(c_1, c_2), \cdots, q(c_{L-2},c_{L-1}), q(c_{L-1},c_0)$. Therefore, the  net result is a  protocol that converges to $Q$  through cyclic satisfaction of $ q(c_0,c_1), q(c_1, c_2), \cdots, q(c_{L-2},c_{L-1}), q(c_{L-1},c_0)$ under strong fairness. This is a contradiction with Theorem   \ref{Qcoloring}. 
\end{proof}

\section{Decidability of Synthesis}
\label{sec:dec}

This section proves the decidability of synthesizing symmetric uni-ring protocols that satisfy the leads-to property $(R \leadsto Q)$,  where $R$ and $Q$ denote non-empty conjunctive state predicates.  We first establish a relation between self-stabilization and leads-to by the following lemma:


 
 \begin{lemma}
\label{decEventualLem}
Let $R$ and $Q$ be conjunctive state predicates specified on uni-rings. There is a symmetric protocol that satisfies $(R  \leadsto Q)$ for unbounded (but finite) ring sizes {\it iff}  there is a symmetric protocol that  stabilizes to  $Q$ for unbounded (but finite) ring sizes. 
\end{lemma}
 
\begin{proof}
Any protocol $p$ that is self-stabilizing to $Q$ ensures convergence to $Q$; i.e., $(true \leadsto Q)$. Since $R \subset true$, it follows that protocol $p$ satisfies $(R  \leadsto Q)$. Now, if there is no protocol that  stabilizes to $Q$, then Theorem \ref{decSynSS} implies that there is no value $\gamma$ in the domain of $x_i$ for which $q(\gamma,\gamma)$ holds. Thus, for a protocol $p'$ to satisfy $(R  \leadsto Q)$, $p'$ must converge to states where $Q$ holds through cyclic satisfaction of $ q(c_0,c_1), q(c_1, c_2), \cdots, q(c_{L-2},c_{L-1}), q(c_{L-1},c_0)$, which is impossible due to Theorem \ref{undecEventual}. 
\end{proof}


\begin{theorem}
\label{decEventual}
 Synthesizing a  symmetric protocol  on uni-rings that  satisfies $(R \leadsto Q)$ (for unbounded ring sizes) is decidable.
 \end{theorem}



 
 To prove Theorem \ref{decEventual}, we present a synthesis algorithm that takes two non-empty and disjoint conjunctive state predicates $R$ and $Q$, and generates a parameterized protocol that satisfies $R \leadsto Q$ on symmetric uni-rings, for unbounded ring sizes. Wlog, we assume that $R \cap Q = \emptyset$; even if $R$ and $Q$ intersect, $i.e., (R \cap Q \neq \emptyset)$ , the synthesis problem is formulated for $(R-X) \leadsto (Q-X)$, where $X=R \cap Q $. (Note that $R \leadsto Q$ is vacuously true in $X$.) We later show that Algorithm \ref{alg:SynLeadsTo} is sound and complete.

\begin{figure}
\vspace*{-5mm}
\caption{Synthesis algorithm for LeadsTo in symmetric uni-rings}
\centering
\fbox{\parbox{1\linewidth}{
\begin{algorithm}{SynLeadsTo($r(x_{i-1},x_i), q(x_{i-1},x_i)$: state predicate, $M$: domain size of $x_i$)}

\label{alg:SynLeadsTo}
\begin{algorithmic}[1]


\STATE Create the locality graphs $G_Q=(V_Q,A_Q)$ and $G_R=(V_R,A_R)$ respectively  for both $q(x_{i-1},x_i)$ and $r(x_{i-1},x_i)$.

\STATE Find a $\gamma \in V_Q$  such that $q(\gamma,\gamma)$ holds and $\gamma$ has not been used before. If no such $\gamma$ exists, then declare that {\sf no solution exists} and {\bf exit}.

\STATE Induce a subgraph $G'_Q=(V'_Q, A'_Q)$ that contains all arcs of $A_Q$ that participate in simple cycles involving $\gamma$. If there is no such subgraph, then $V'_Q = \{ \gamma \}$ and  $A'_Q = \emptyset$.

\STATE Compute a spanning tree $\tau$ of $G'_Q$ rooted at $\gamma$.

\STATE Let $V'_R$ be the subset of $V_R$ that do not participate in any cycle. 
 
\STATE Let $V'_{Rleaf}$ be the set of vertices $v \in V'_R$ that are leaves in $\tau$. Remove the outgoing arc of each $v \in V'_{Rleaf}$, hence creating a tree $\tau'$ (which is no longer a {\em spanning} tree of $G'_Q$).

\STATE For each node $v \in (V_Q - (V'_Q \cup V'_{Rleaf}))$, include an arc from $v$ to the root $\gamma$ of the spanning tree $\tau'$ of $G'_Q$, unless $r(v,\gamma)$ holds.

\STATE For each node $v \in V_Q$ where $r(v,\gamma)$ holds, include an arc $(v,l)$, where $l$ is a leaf in $\tau'$.  The resulting graph would still be a tree, denoted  $\tau''$. Include a self-loop $(\gamma,\gamma)$ at the root of  $\tau''$. If $\tau''$ has no leaves in common with any cycle in $G_R$, then go to Step 2.

\STATE For each leaf vertex $a$ in $\tau''$, label its outgoing arc $(a,c)$ with a value $b \in \IntMod{M}$ iff  $b \neq c$  and $r(a,b) \wedge \neg q(a,b)$.

\STATE  For any other arc $(a,c)$ in $\tau''$, label it with a value $b \in \IntMod{M}$ iff $b \neq c$ and $\neg q(a,b)$.

\STATE For each labelled arc $(a,b,c)$ in $\tau''$ (where $b$ is the label of arc $(a,c)$), generate a parameterized action $x_{i-1} = a \wedge x_i = b \rightarrow x_i := c$.

\end{algorithmic}
\end{algorithm}}}
\end{figure}


For a specific $\gamma$, Algorithm \ref{alg:SynLeadsTo} performs Steps 3 to 8 in order to determine if there is a solution for that $\gamma$. In Steps 3 to 8, Algorithm \ref{alg:SynLeadsTo} constructs the underlying structure of an action graph that will form the synthesized protocol $p$ after the labeling steps of 9 and 10. A correct protocol must meet certain constraints that are the minimum requirements for a protocol $p$ that satisfies $R \leadsto Q$. For instance, $p$ should only guarantee $R \leadsto Q$ starting in $R$. That is why Step 5 calculates the set of values in $\IntMod{M}$ that do not appear in any state of $R$, and Step 6 removes them from the spanning tree of $G'_Q$.  Moreover, since we assume $R \cap Q = \emptyset$, the cycles of $G_R$ and $G_Q$ are arc-disjoint (due to Theorem \ref{thm:cycleGState}). Thus, any correct protocol cannot include an arc $(a,c) \in A_R$ for which $r(a,c)$ holds and $c = \gamma$. Otherwise, the processes will include an action $x_{i-1}=a \wedge x_i = b \rightarrow x_i :=c$, for some $b$ where $\neg q(a,b)$. Such an action would set the locality of each process to a state (i.e., $x_{i-1}=a \wedge x_i = c$) where $R$ holds, thereby preventing reachability to $Q$. That is why Step 7 excludes such arcs. After eliminating such vertices and arcs through Steps 3 to 8, if we have a tree-like structure that has no leaves in common with the cycles in $G_R$, then we move to Step 2 and repeat the same process for a different $\gamma$. Algorithm \ref{alg:SynLeadsTo} exits only if no valid action graph can be built for any $\gamma$. 



\begin{theorem}[Soundness]
Algorithm \ref{alg:SynLeadsTo} is sound. 
\end{theorem}

\begin{proof}
We show that if Algorithm \ref{alg:SynLeadsTo} generates a parameterized protocol $p$ for two disjoint predicates $R$ and $Q$, then $p$ satisfies $R \leadsto Q$ for unbounded ring sizes. To prove this, we show that any computation of $p$ starting in a state $s \in R$ will be deadlock-free and livelock-free outside $Q$ and will eventually reach a state $s_f$ in $Q$.

{\it Deadlock-freedom}: Since the synthesized action graph, denoted $AG$, is a tree with a self-loop on its root $\gamma$, $AG$ must have some leaves. Step 8 of the algorithm ensures that any vertex $v \neq \gamma$ that participate in a cycle in the locality graph $G_R$ of the state predicate $R$ appears as a leaf in the tree. 
The labeling method of Step 9 guarantees that any leaf $a$ of $AG$ has an outgoing arc $(a,c)$ (to some vertex $c \in \IntMod{M}$) with a label $b$ such that $r(a,b)$ holds. Step 11 would translate each labeled arcs $(a,b,c)$  to a parameterized action $x_{i-1}=a \wedge x_i = b \rightarrow x:=c$. By Theorem \ref{thm:cycleGState}, we also know that cycles in locality graphs characterize global states of uni-rings. Thus, the guard condition of such an action evaluates to {\it true}  in a state in $R$ because $r(a,b)$ holds. Thus, starting in $R$, there is at least one enabled action; i.e., deadlock-freedom in $R$. We now show that the computations of the synthesized protocol $p$ remain deadlock-free until reaching a state in $Q$. The only values in $\IntMod{M}$ that are excluded from the synthesized action graph $AG$ include those values that do not participate in any cycle in $G_R$; i.e., those values do not appear in states in $R$. If computations of $p$ reach a state outside $R \vee Q$, the labeling method of Step 10 ensures that there is some enabled action; hence deadlock-freedom.

{\it Livelock-freedom}: The only type of propagation included in the synthesized action graph is $(\gamma, b, \gamma)$. Thus, there are no propagations that lead each other circularly. That is, based on Theorem \ref{thm:LivelockIFF}, the synthesized protocol is livelock-free.

{\it Reachability of $Q$}: By deadlock-freedom and livelock-freedom,  each process will eventually satisfy $q(\gamma,\gamma)$, which would result in a global state of the ring in $Q$.
\end{proof}

\begin{theorem}[Completeness]
Algorithm \ref{alg:SynLeadsTo} is  complete. 
\end{theorem}

\begin{proof}
Let $p$ be a parameterized protocol that satisfies $R \leadsto Q$  but Algorithm \ref{alg:SynLeadsTo} fails to generate $p$. 
By Theorem \ref{undecEventual}, $p$ cannot satisfy $R \leadsto Q$  through cyclic satisfaction of some values; i.e., the action graph of $p$, denoted $A_p$,  can include only cycles of length 1; i.e., self-loops. Further, if a node $v$ has a self-loop in $A_p$, then $v$ cannot have any other outgoing labeled arc; otherwise, $A_p$ will include either  cycles of length greater than one, which contradicts Theorem \ref{undecEventual}, or ends in nodes without any outgoing arcs; i.e., deadlock. Thus, there must be a value $\gamma \in \IntMod{M}$  for which $q(\gamma,\gamma)$ holds and there is some value $v \in  \IntMod{M}$, where $v \neq \gamma $ and $q(v, \gamma)$ holds too. That is,  $A_p$ must include  $\gamma$ as well as another labeled arc $(v, b, \gamma)$ for some $b \in \IntMod{M}$, where $v \neq \gamma$ and $b \neq \gamma$. This means that Algorithm \ref{alg:SynLeadsTo} would actually find such $\gamma$ in Step 2. 
 The  action graph  $A_p$  must also include some source nodes without any incoming arcs. Otherwise, it would include cycles of length greater than one (which again contradicts Theorem \ref{undecEventual}). Such source nodes must also intersect with the vertices of some cycles in $V_R$ and have outgoing labeled arcs $(a,b,c)$ such that $r(a,b)$ holds and $a$ is a vertex in a cycle in $G_R$. Otherwise, $p$ deadlocks in $R$, which contradicts with $p$  satisfying $R \leadsto Q$. As such, Algorithm \ref{alg:SynLeadsTo} would have included such labeled arcs in Steps 8 and 9, and would have created the action graph $A_p$ sinking towards $\gamma$.
\end{proof}

\begin{theorem}
Algorithm \ref{alg:SynLeadsTo} has an asymptotic polynomial time complexity in the domain size $M$.
\end{theorem}

\begin{proof}
Other than Step 3, it is trivial to see that all the other steps would take polynomial amount of time (in $M=|V_Q|=$DomainSize$(x_i)$) to compute. For Step 3, we first remove self-loops. Then, we start eliminating any vertex $v \neq \gamma$ in the locality graph $G_Q = (V_Q,A_Q)$ that has no outgoing arcs. This would result in removing the incoming arcs of $v$ too. We continue such vertex/arc removals until all remainig vertices have at least one outgoing arc, or no vertex remains. The remaining sub-graph that contains $\gamma$ would be used for subsequent steps of the algorithm. This process takes $O(M)$. Therefore, the asymptotic time complexity of Algorithm \ref{alg:SynLeadsTo} is polynomial in $M$.
\end{proof}






\begin{theorem}
\label{decDisjEventual}
Let $R_i$ and $Q$ be conjunctive state predicates, for $1 \leq i \leq k$ and $k>1$. Synthesizing a  parameterized protocol  on symmetric uni-rings that  satisfies $((R_1 \vee R_2 \cdots \vee R_k) \leadsto Q)$ is decidable. 
\end{theorem}


\begin{proof}
One can apply Algorithm \ref{alg:SynLeadsTo} $k$ times for each $R_i$ and decide if there is some $1 \leq i \leq k$ for which there is a protocol that satisfies $R_i \leadsto Q$. Any such protocol is also a solution for $((R_1 \vee R_2 \cdots \vee R_k) \leadsto Q)$.
\end{proof}

\begin{theorem}
\label{decConjEventual}
Let $R_i$ be conjunctive state predicates, for $1 \leq i \leq k$ and $k>1$. Synthesizing a  parameterized protocol  on symmetric uni-rings that  satisfies $((R_1 \wedge R_2 \cdots \wedge R_k) \leadsto Q)$ is decidable. 
\end{theorem}


\begin{proof}
Since $R_i$ predicates are conjunctive, we apply Algorithm \ref{alg:SynLeadsTo} while  identifying  cycles in the intersection of the locality graphs of $R_i$, for $1 \leq i \leq k$. The rest of Algorithm \ref{alg:SynLeadsTo} remains the same.
\end{proof}

\begin{theorem}
\label{decEventualTwoQsConj}
Let  $R, Q_1$ and $Q_2$ be conjunctive state predicates and $Q_1 \cap Q_2 \neq \emptyset$.  Synthesizing a  parameterized protocol  on symmetric uni-rings that  satisfies $(R \leadsto (Q_1 \wedge Q_2))$ is decidable.
\end{theorem}

\begin{proof}
We use Algorithm \ref{alg:SynLeadsTo} by finding a value $\gamma$ in the domain of the variable $x_i$ (of the template process $P_i$) such that $ q_1(\gamma,\gamma) \wedge q_2(\gamma,\gamma)$ holds. Moreover, it is straightforward to algorithmically find a common cycle in the locality graphs of $Q_1$ and $Q_2$ that contains $\gamma$ (e.g., by running a DFS algorithm starting at $\gamma$). There is a parameterized protocol that satisfies $(R \leadsto (Q_1 \wedge Q_2))$ {\em iff} Algorithm \ref{alg:SynLeadsTo}  generates a protocol (due to its   completeness).
\end{proof}

\begin{theorem}
\label{decEventualTwoQsDisj}
Let  $R, Q_1$ and $Q_2$ be conjunctive state predicates.  Synthesizing a  parameterized protocol  on symmetric uni-rings that  satisfies $(R \leadsto (Q_1 \vee Q_2))$ is decidable.
\end{theorem}

\begin{proof}
It is straightforward to see that a protocol $p$ satisfies $(R \leadsto (Q_1 \vee Q_2))$ {\em iff} $p$ satisfies  $(R \leadsto Q_1)$ or $(R \leadsto Q_2)$. 
We execute Algorithm \ref{alg:SynLeadsTo} once for $R \leadsto Q_1$ and another time for $R \leadsto Q_2$ in order to decide if a solution for $(R \leadsto (Q_1 \vee Q_2))$  exists.
\end{proof}

\section{Case Studies}
\label{sec:cases}

This section presents four case studies of using Algorithm \ref{alg:SynLeadsTo} for the synthesis of symmetric uni-rings. Section \ref{sumnot2} discusses the synthesis of the Sum-Not-2 protocol, and Section \ref{sec:sumTwo} presents the dual of Sum-Not-2 protocol.  Section \ref{sec:parity} illustrates the synthesis of a symmetric uni-ring for solving the parity problem in distributed computing. Section \ref{sec:agree} presents the synthesis of an agreement protocol on uni-rings. To increase our confidence in the proposed synthesis method, we have model checked all protocols synthesized in this section using SPIN \cite{spin97} up to the extent  our computational resources permit. The Promela models are available at \url{http://asd.cs.mtu.edu/projects/ProTop/index.html}.


\subsection{Sum-Not-2}
\label{sumnot2}
The Sum-Not-2 protocol is a simple but non-trivial example that can clearly demonstrate the complexity of synthesizing parameterized uni-rings that satisfy leadsto properties. In this protocol, we specify the set of initial states for even-size uni-rings where the summation of $x_{i-1}$ and $x_i$ is equal to two (modulo $M=4$) but $x_i \neq 1$. That is, $R = \forall  i \in \IntMod{N}: r(x_{i-1}, x_i)$, where $ r(x_{i-1}, x_i) \nameq  (x_i = 2 \wedge x_{i-1}=0) \vee (x_i = 0 \wedge x_{i-1}=2)$, and  $N$ denotes the number of processes (i.e., ring size). The objective is to synthesize a protocol that eventually reaches states where the summation of $x_{i-1}$ and $x_i$ is no longer equal to two (for all processes), and it is not the case that all processes have a value of zero; i.e., $Q = \forall  i \in  \IntMod{N}: q(x_{i-1},x_i)$, where $q(x_{i-1},x_i) \equiv ((x_{i-1} \oplus_4 x_i) \neq 2) \wedge ( (x_{i-1}  \neq 0) \vee (x_i \neq 0)) $, and  $\oplus_4$ denotes addition modulo $M=4$. We require that $R \leadsto Q$ is satisfied from $R$ for all even values of $N$. 


\noindent{\it Step 1}:\ As the first step of Algorithm  \ref{alg:SynLeadsTo}, we construct the locality graphs of $R$ and $Q$ illustrated in Figure \ref{fig:locNot2}. Each arc $(a,b)$ captures the fact that $r(a,b)$ (respectively, $q(a,b)$) holds. For example, the only arcs in  Figure \ref{fig:locNot2}-(a) are between 0 and 2 because $r(x_{i-1},x_i)$ does not hold for any other pairs of values in $\IntMod{4}$. As another example, Figure \ref{fig:locNot2}-(b) lacks any arcs between 0 and 2 because their summation adds up to 2, which violates $Q$. Moreover, the only vertex that has a self-loop is 2 because the arc $(0,0)$ violates the second conjunct of $Q$ and arcs $(1,1)$ and $(3,3)$ violate the first conjunct of $Q$ (i.e., $1 \oplus_4 1 = 2$ and $3 \oplus_4 3 = 2$).

\renewcommand{\TikzDecSyntInstance}{%
\tikzstyle{snod}=[draw,circle]
\tikzstyle{phantnod}=[inner sep=0pt]
\draw
( 0, 0) node[snod] (0) {$0$}
( 1, 0) node[snod] (1) {$1$}
( 0,-1) node[snod] (2) {$2$}
( 1,-1) node[snod] (3) {$3$}
;
\tikzstyle{lab}=[font=\scriptsize]
\tikzstyle{xn}=[draw,arrows=-latex,font=\scriptsize]}

\begin{figure}[h]
\centering
\subfloat[Locality graph $G_R$ representing predicate $R$.]{
\makebox[.2\textwidth]{%
\begin{tikzpicture}[x=1cm,y=1.2cm]
\TikzDecSyntInstance
\draw[xn]
(0) edge[bend  left=15] (2)
(2) edge[bend  left=15] (0)
;
\end{tikzpicture}%
\label{fig:locRNot2}
}}%
\hspace*{35mm}
\subfloat[Locality graph $G_Q$ representing predicate $Q$.]{
\makebox[.2\textwidth]{%
\begin{tikzpicture}[x=1.5cm,y=1.2cm]
\TikzDecSyntInstance
\draw[xn]
(2) edge[out=200,in=160,looseness=7]  (2)
(0) edge[bend  left=15] (1)
(1) edge[bend  left=15] (0)
(3) edge[bend  left=15] (1)
(1) edge[bend  left=15] (3)
(1) edge[bend  left=15] (2)
(2) edge[bend  left=15] (1)
(2) edge[bend  left=15] (3)
(3) edge[bend  left=15] (2)
(0) edge[bend  left=15] (3)
(3) edge[bend  left=15] (0)
;
\end{tikzpicture}%
\label{fig:locQNot2}
}}%
\caption{Locality graphs for the predicates $R$ and $Q$ in SumNotTwo.}
\label{fig:locNot2}
\end{figure}
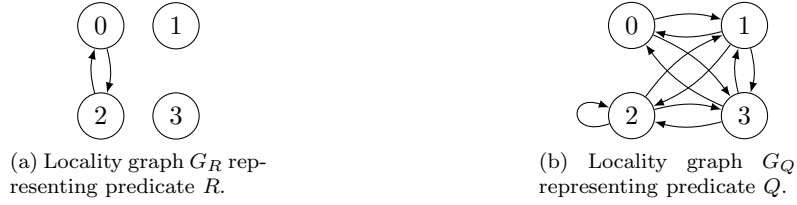

\noindent{\it Step 2}:\ Since $q(2,2)$ holds in $G_Q$ (see Figure \ref{fig:locNot2}-(b)), we set $\gamma$ to 2.

\noindent{\it Step 3}:\ Figure  \ref{fig:treeNot2}-(a) illustrates the induced subgraph $G'_Q$ including simple cycles of $G_Q$ that contain $\gamma$  (e.g., the simple cycle $(0, 3, 2, 1, 0)$).

\noindent{\it Step 4}:\    Figure \ref{fig:treeNot2}-(b) illustrates a  spanning tree $\tau$ rooted at $\gamma = 2$, where each arc $(a,b)$ denotes that $b$ is the parent of $a$ in $\tau$. Notice that, there may be several spanning trees; i.e., the solution is not unique.

\renewcommand{\TikzDecSyntInstance}{%
\tikzstyle{snod}=[draw,circle]
\tikzstyle{phantnod}=[inner sep=0pt]
\draw
( 0, 0) node[snod] (0) {$0$}
( 1, 0) node[snod] (1) {$1$}
( 0,-1) node[snod] (2) {$2$}
( 1,-1) node[snod] (3) {$3$}
;
\tikzstyle{lab}=[font=\scriptsize]
\tikzstyle{xn}=[draw,arrows=-latex,font=\scriptsize]}

\begin{figure}[h]
\centering
\subfloat[Induced subgraph $G'_Q$.]{
\makebox[.2\textwidth]{%
\begin{tikzpicture}[x=1.5cm,y=1.2cm]
\TikzDecSyntInstance
\draw[xn]
(0) edge[bend  left=15] (1)
(1) edge[bend  left=15] (0)
(3) edge[bend  left=15] (1)
(1) edge[bend  left=15] (3)
(1) edge[bend  left=15] (2)
(2) edge[bend  left=15] (1)
(2) edge[bend  left=15] (3)
(3) edge[bend  left=15] (2)
(0) edge[bend  left=15] (3)
(3) edge[bend  left=15] (0)
;
\end{tikzpicture}%
\label{fig:locQNot2}
}}%
\hspace*{35mm}
\subfloat[Spanning tree $\tau$ of the induced subgraph $G'_Q$ with the root $\gamma = 2$.]{
\makebox[.2\textwidth]{%
\begin{tikzpicture}[x=2cm,y=1.5cm]
\TikzDecSyntInstance
\tikzstyle{pruned}=[dashed]
\draw[xn]
(1) edge[bend  left=15, pruned] (3)
(3) edge[bend  left=15] (2)
(0) edge[bend  left=15] (3)
;
\end{tikzpicture}%
\label{fig:treeQNot2}
}}%
\caption{Construction of the spanning tree of SumNotTwo.}
\label{fig:treeNot2}
\end{figure}
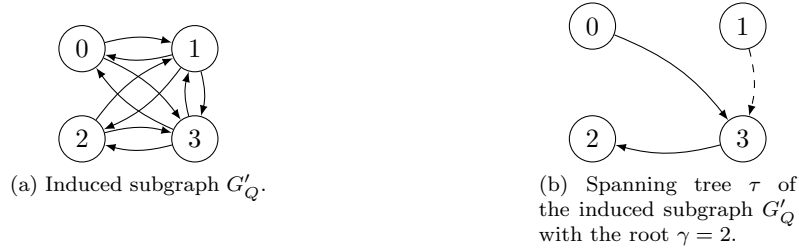

\noindent{\it Step 5}:\ Figure \ref{fig:locNot2}-(a) illustrates that vertices 1 and 3 do not participate in any cycles; hence $V'_R = \{1, 3\}$.

\noindent{\it Step 6}:\ Since $1$ is the only vertex of $V'_R$ that is also a leaf of the spanning tree $\tau$, we eliminate its outgoing arc to $3$, illustrated by the dashed arrow in Figure \ref{fig:treeNot2}-(b).

\noindent{\it Step 7}:\  Since $V'_Q = V_Q$, Step 7 does not make any changes.


\noindent{\it Step 8}:\ The only leaf of the tree  $\tau'$  for which $r(0, 2)$ holds is $0$. Thus, we include just the self-loop $(2,2)$ to generate the tree $\tau''$  in Figure  \ref{fig:synNot2}-(a).

\renewcommand{\TikzDecSyntInstance}{%
\tikzstyle{snod}=[draw,circle]
\tikzstyle{phantnod}=[inner sep=0pt]
\draw
( 0, 0) node[snod] (0) {$0$}
( 1, 0) node[snod] (1) {$1$}
( 0,-1) node[snod] (2) {$2$}
( 1,-1) node[snod] (3) {$3$}
;
\tikzstyle{lab}=[font=\scriptsize]
\tikzstyle{xn}=[draw,arrows=-latex,font=\scriptsize]}

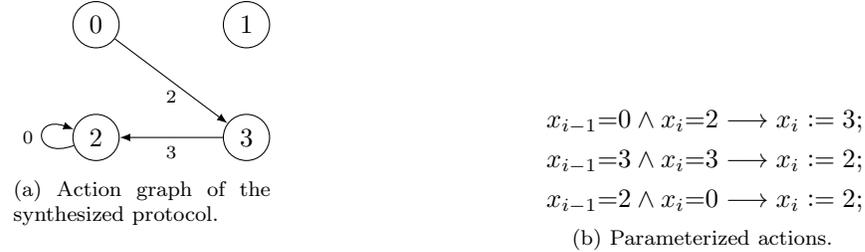
\begin{figure}[h]
\hspace*{30mm}
\subfloat[Action graph of the synthesized protocol.]{
\makebox[.2\textwidth]{%
\begin{tikzpicture}[x=2cm,y=1.5cm]
\TikzDecSyntInstance
\tikzstyle{pruned}=[dashed]
\draw[xn]
(2) edge [out=200,in=160,looseness=7]  node[left]  {0} (2)
(0) edge node [below] {2}(3)
(3) edge node[below] {3}(2)
;
\end{tikzpicture}%
\label{fig:synNot2Act}
}}%
\hspace*{-10mm}
\subfloat[Parameterized actions.]{
\makebox[.8\textwidth]{%
$\begin{alignedat}{4}%
  &x_{i-1}{=}0 &{}\land{}& x_i{=}2 &{}\acteq{}& x_i \gets 3;
\\&x_{i-1}{=}3 &{}\land{}& x_i{=}3 &{}\acteq{}& x_i \gets 2;
\\&x_{i-1}{=}2 &{}\land{}& x_i{=}0 &{}\acteq{}& x_i \gets 2;
\end{alignedat}$
\label{fig:SumNotTwoSynt:acts}
}}%
\caption{Synthesized parameterized protocol that satisfies $R \leadsto Q$.}
\label{fig:synNot2}
\end{figure}

\noindent{\it Steps 9 and 10}:\ Consider the arc $(3,2)$. The candidate labels of this arc include $0, 1$ and $3$. We exclude $2$ because it is equal to the parent vertex of $3$. Since $q(3,0)$ and $q(3,1)$ are {\it true} and $q(3,3)$ is {\it false}, the only acceptable label for the arc $(3,2)$ is $3$ (see Figure \ref{fig:synNot2}-(a)). The arc $(0, 3)$ has the label $2$ because $r(0,2) \wedge \neg q(0,2)$ holds. Likewise, the self-loop on $2$ gets $0$ as its label.

\noindent{\it Step 11}:\ Figure \ref{fig:synNot2}-(b) illustrates the synthesized parameterized actions for any even-size uni-ring that satisfies $R \leadsto Q$.

\subsection{SumTwo Protocol}
\label{sec:sumTwo}
We consider the dual of the previous example, which provides another interesting case. Let $R$ be $ \forall  i \in \IntMod{N}: (x_i  \oplus_4 x_{i-1} \neq 2) \wedge (x_i \oplus_4 1 =  x_{i-1})$, and $Q$ be $\forall  i \in  \IntMod{N}: ((x_{i-1} \oplus_4 x_i) = 2) \wedge ((x_{i-1} \neq 1) \vee (x_i \neq 1))$, where  $N$ denotes the number of processes (i.e., ring size), and $\oplus_4$ represents addition modulo $M=4$. (Addition and subtraction in subscripts are done modulo $N$.)
We would like to synthesize a protocol that eventually reaches states where the summation of $x_{i-1}$ and $x_i$ is equal to two (for all processes), and it is not the case that all processes have a value $1$. We require that $R \leadsto Q$ is satisfied from $R$ for all even values of $N$.

\noindent{\it Step 1}: \ Figure \ref{fig:loc2} illustrates the locality graphs of the predicates $R$ and $Q$.

\renewcommand{\TikzDecSyntInstance}{%
\tikzstyle{snod}=[draw,circle]
\tikzstyle{phantnod}=[inner sep=0pt]
\draw
( 0, 0) node[snod] (0) {$0$}
( 1, 0) node[snod] (1) {$1$}
( 0,-1) node[snod] (2) {$2$}
( 1,-1) node[snod] (3) {$3$}
;
\tikzstyle{lab}=[font=\scriptsize]
\tikzstyle{xn}=[draw,arrows=-latex,font=\scriptsize]}

\begin{figure}[h]
\centering
\subfloat[Locality graph $G_R$ representing predicate $R$.]{
\makebox[.2\textwidth]{%
\begin{tikzpicture}[x=1.6cm,y=1.2cm]
\TikzDecSyntInstance
\draw[xn]
(0) edge[bend  left=15] (3)
(3) edge[bend  left=15] (2)
(1) edge[bend  left=-15] (0)
(2) edge[bend  left=15] (1)
;
\end{tikzpicture}%
\label{fig:locR2}
}}%
\hspace*{35mm}
\subfloat[Locality graph $G_Q$ representing predicate $Q$.]{
\makebox[.2\textwidth]{%
\begin{tikzpicture}[x=1.5cm,y=1.2cm]
\TikzDecSyntInstance
\draw[xn]
(3) edge[out=200,in=160,looseness=7]  (3)
(0) edge[bend  left=15] (2)
(2) edge[bend  left=15] (0)
;
\end{tikzpicture}%
\label{fig:locQ2}
}}%
\caption{Locality graphs of SumTwo protocol.}
\label{fig:loc2}
\end{figure}
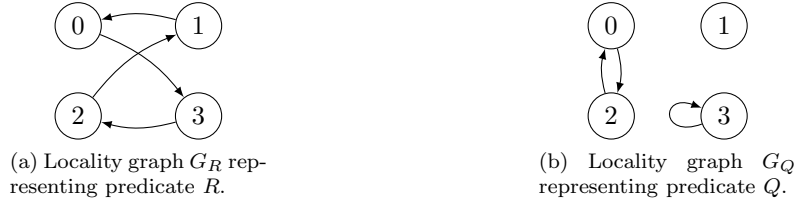

\noindent{\it Steps 2, 3, 4}: \  Since $q(3,3)$, we have $\gamma = 3$ (see Figure \ref{fig:loc2}-(b)). The induced subgraph $G'_Q$ includes the vertex $3$. Thus, the resulting spanning tree would include a single vertex; i.e., $3$. 

\noindent{\it Steps 5, 6}: \ Since all vertices of $G_R$ participate in a cycle (Figure \ref{fig:loc2}-(a)), we have $V'_R = \emptyset$. As such, there is nothing to be done in Step 6.

\noindent{\it Steps 7 and 8}: \ Including arcs from vertices that are in $V_Q - V'_Q$ to $\gamma$ would result in the spanning tree $\tau$ in Figure \ref{fig:loc2Tree}-(a).  Since $r(0,3)$ holds, $\tau'$ would look like the tree in  Figure \ref{fig:loc2Tree}-(b) excluding the dashed arc and the self-loop.  
 Including an arc from 0 to the leaf 1 as well as the self-loop $(3,3)$ would generate the tree $\tau''$ in Figure \ref{fig:loc2Tree}-(b).

\renewcommand{\TikzDecSyntInstance}{%
\tikzstyle{snod}=[draw,circle]
\tikzstyle{phantnod}=[inner sep=0pt]
\draw
( 0, 0) node[snod] (0) {$0$}
( 1, 0) node[snod] (1) {$1$}
( 0,-1) node[snod] (2) {$2$}
( 1,-1) node[snod] (3) {$3$}
;
\tikzstyle{lab}=[font=\scriptsize]
\tikzstyle{xn}=[draw,arrows=-latex,font=\scriptsize]}

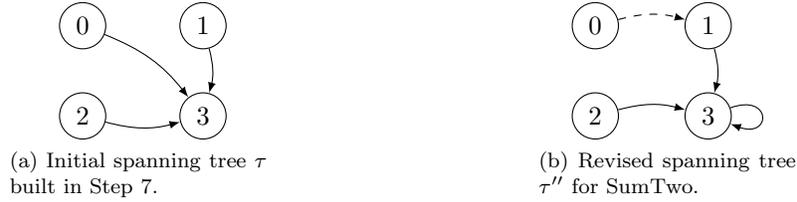
\begin{figure}[h]
\centering
\subfloat[Initial spanning tree $\tau$ built in Step 7.]{
\makebox[.2\textwidth]{%
\begin{tikzpicture}[x=1.6cm,y=1.2cm]
\TikzDecSyntInstance
\draw[xn]
(0) edge[bend  left=15] (3)
(1) edge[bend  left=15] (3)
(2) edge[bend  left=-15] (3)
;
\end{tikzpicture}%
}}%
\hspace*{35mm}
\subfloat[Revised spanning tree $\tau''$ for SumTwo.]{
\makebox[.2\textwidth]{%
\begin{tikzpicture}[x=1.5cm,y=1.2cm]
\TikzDecSyntInstance
\tikzstyle{pruned}=[dashed]
\draw[xn]
(3) edge[out=20,in=-20,looseness=7]  (3)
(0) edge[bend  left=15, pruned] (1)
(1) edge[bend  left=15] (3)
(2) edge[bend  left=15] (3)
;
\end{tikzpicture}%
}}%
\caption{Spanning tree built for the SumTwo protocol.}
\label{fig:loc2Tree}
\end{figure}

\noindent{\it Steps 9, 10, 11}:\ The remaining steps of Algorithm \ref{alg:SynLeadsTo} would generate the action graph of Figure \ref{fig:syn2}-(a), which would be translated to the parameterized actions of the synthesized SumTwo protocol in Figure  \ref{fig:syn2}-(b).

\renewcommand{\TikzDecSyntInstance}{%
\tikzstyle{snod}=[draw,circle]
\tikzstyle{phantnod}=[inner sep=0pt]
\draw
( 0, 0) node[snod] (0) {$0$}
( 1, 0) node[snod] (1) {$1$}
( 0,-1) node[snod] (2) {$2$}
( 1,-1) node[snod] (3) {$3$}
;
\tikzstyle{lab}=[font=\scriptsize]
\tikzstyle{xn}=[draw,arrows=-latex,font=\scriptsize]}

\begin{figure}[h]
\hspace*{35mm}
\subfloat[Action graph of the synthesized protocol.]{
\makebox[.2\textwidth]{%
\begin{tikzpicture}[x=2cm,y=1.5cm]
\TikzDecSyntInstance
\tikzstyle{pruned}=[dashed]
\draw[xn]
(3) edge [out=20,in=-20,looseness=7]  node[right]  {$0|1|2$} (3)
(0) edge node [below] {3}(1)
(1) edge node[right] {$0|2$}(3)
(2) edge node[below] {1}(3)
;
\end{tikzpicture}%
}}%
\hspace*{-18mm}
\subfloat[Parameterized actions.]{
\makebox[.8\textwidth]{%
$\begin{alignedat}{4}%
  &x_{i-1}{=}0 &{}\land{}& x_i{=}3 &{}&{}&{}\acteq{}& x_i \gets 1;
\\&x_{i-1}{=}1 &{}\land{}& (x_i{=}0 &{}\lor{}x_i{=}2)&{}&{}\acteq{}& x_i \gets 3;
\\&x_{i-1}{=}2 &{}\land{}& x_i{=}1 &{}&{}&{}\acteq{}& x_i \gets 3;
\\&x_{i-1}{=}3 &{}\land{}& (x_i{=}0&{}\lor{}x_i{=}1&{}\lor{}x_i{=}2) &{}\acteq{}& x_i \gets 3;
\end{alignedat}$
}}%
\caption{Synthesized parameterized SumTwo protocol that satisfies $R \leadsto Q$.}
\label{fig:syn2}
\end{figure}
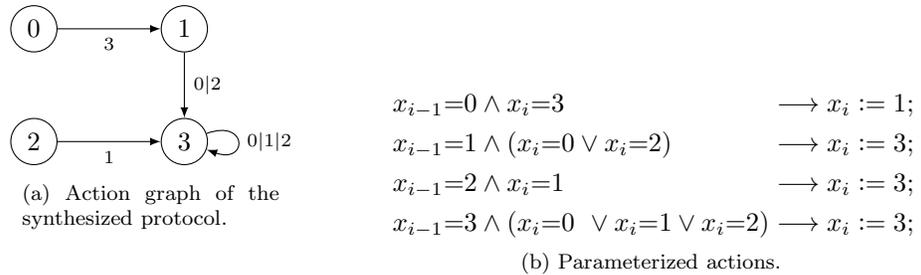

\subsection{Parity Protocol}
\label{sec:parity}

The {\it Parity} protocol solves the problem of identifying a common parity amongst the nodes of a distributed system without a central coordinator dictating what the parity should be. In this case study, we synthesize a parameterized protocol for symmetric uni-rings that ensures reachability to an even parity in the entire ring from states where there is an odd parity. In fact, the synthesized protocol provides an algorithm for switching from odd to even parity amongst the nodes of a uni-ring. More precisely, we would like the protocol to satisfy $R \leadsto Q$, where $R = \forall i \in \IntMod{N}: (((x_i \ominus_4 x_{i-1}) \mod 2= 0) \wedge (x_i \mod 2 \neq 0))$,  $Q = \forall i \in \IntMod{N}: (((x_i \ominus_4 x_{i-1}) \mod 2= 0) \wedge (x_i \mod 2 =0))$,  $M=4$ and $\ominus_4$ denotes subtraction modulo 4.

\noindent{\it Steps 1 and 2}:\ We first construct the locality graphs $G_R$ and $G_Q$ (illustrated in Figure \ref{fig:parity}). Notice that both 0 and 2 can be considered as $\gamma$; we let $\gamma = 2$.


\renewcommand{\TikzDecSyntInstance}{%
\tikzstyle{snod}=[draw,circle]
\tikzstyle{phantnod}=[inner sep=0pt]
\draw
( 0, 0) node[snod] (0) {$0$}
( 1, 0) node[snod] (1) {$1$}
( 0,-1) node[snod] (2) {$2$}
( 1,-1) node[snod] (3) {$3$}
;
\tikzstyle{lab}=[font=\scriptsize]
\tikzstyle{xn}=[draw,arrows=-latex,font=\scriptsize]}

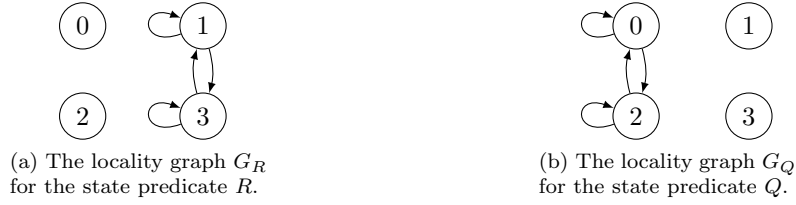
\begin{figure}[h]
\centering
\subfloat[The locality graph $G_R$ for the state predicate $R$.]{
\makebox[.2\textwidth]{%
\begin{tikzpicture}[x=1.6cm,y=1.2cm]
\TikzDecSyntInstance
\draw[xn]
(1) edge[bend  left=15] (3)
(3) edge[bend  left=15] (1)
(1) edge[out=200,in=160,looseness=7]  (1)
(3) edge[out=200,in=160,looseness=7]  (3)
;
\end{tikzpicture}%
}}%
\hspace*{35mm}
\subfloat[The locality graph $G_Q$ for the state predicate $Q$.]{
\makebox[.2\textwidth]{%
\begin{tikzpicture}[x=1.5cm,y=1.2cm]
\TikzDecSyntInstance
\draw[xn]
(0) edge[bend  left=15] (2)
(2) edge[bend  left=15] (0)
(0) edge[out=200,in=160,looseness=7]  (0)
(2) edge[out=200,in=160,looseness=7]  (2)
;
\end{tikzpicture}%
}}%
\caption{Locality graphs of the Parity protocol.}
\label{fig:parity}
\end{figure}

\noindent{\it Steps 3 and 4}:\ We induce the subgraph $G'_Q$ from $G_Q$ by considering the simple cycle between 0 and 2. We then compute the spanning tree of  $G'_Q$ rooted at $\gamma = 2$.

\renewcommand{\TikzDecSyntInstance}{%
\tikzstyle{snod}=[draw,circle]
\tikzstyle{phantnod}=[inner sep=0pt]
\draw
( 0, 0) node[snod] (0) {$0$}
( 1, 0) node[snod] (1) {$1$}
( 0,-1) node[snod] (2) {$2$}
( 1,-1) node[snod] (3) {$3$}
;
\tikzstyle{lab}=[font=\scriptsize]
\tikzstyle{xn}=[draw,arrows=-latex,font=\scriptsize]}

\begin{figure}[h]
\centering
\subfloat[The induced sub-graph $G'_Q$.]{
\makebox[.2\textwidth]{%
\begin{tikzpicture}[x=1.6cm,y=1.2cm]
\TikzDecSyntInstance
\draw[xn]
(0) edge[bend  left=15] (2)
(2) edge[bend  left=15] (0)
;
\end{tikzpicture}%
}}%
\hspace*{35mm}
\subfloat[The spanning tree $\tau$ of $G'_Q$ rooted at $\gamma$.]{
\makebox[.2\textwidth]{%
\begin{tikzpicture}[x=1.5cm,y=1.2cm]
\TikzDecSyntInstance
\draw[xn]
(0) edge[bend  left=15] (2)
;
\end{tikzpicture}%
}}%
\caption{Induced subgraph and its spanning tree $\tau$ of the Parity protocol.}
\label{fig:parityTree}
\end{figure}
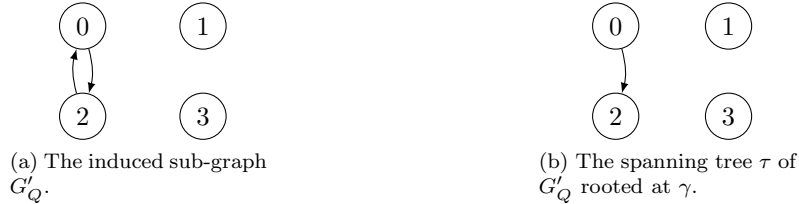

\noindent{\it Steps 5, 6}:\ The set $V'_R$ includes vertices 0 and 2, and the only vertex that is a leaf in $\tau$ is 0; hence $V'_{Rleaf}=\{0\}$. Removing the outgoing arc $(0, 2)$ will create the tree $\tau'$ with the single vertex 2.

\noindent{\it Steps 7 and 8}:\ We now compute $V_Q - (V'_Q \cup V'_{Rleaf})$, which is equal to $\{1, 3\}$. Thus, we include the arcs $(1,2)$ and $(3,2)$ to generate the tree $\tau''$. We also include the self-loop $(2,2)$ in $\tau''$ to generate the structure of the action graph of the synthesized protocol in Figure \ref{fig:synParity}-(a).  Notice that, $r(1,2)$ and $r(3,2)$ do not hold. Since the only leaf in $\tau'$ is vertex 2, arc $(0, 2)$ cannot be included in $\tau''$.

\renewcommand{\TikzDecSyntInstance}{%
\tikzstyle{snod}=[draw,circle]
\tikzstyle{phantnod}=[inner sep=0pt]
\draw
( 0, 0) node[snod] (0) {$0$}
( 1, 0) node[snod] (1) {$1$}
( 0,-1) node[snod] (2) {$2$}
( 1,-1) node[snod] (3) {$3$}
;
\tikzstyle{lab}=[font=\scriptsize]
\tikzstyle{xn}=[draw,arrows=-latex,font=\scriptsize]}

\begin{figure}[h]
\centering
\subfloat[The tree $\tau''$.]{
\makebox[.2\textwidth]{%
\begin{tikzpicture}[x=2cm,y=1.5cm]
\TikzDecSyntInstance
\tikzstyle{pruned}=[dashed]
\draw[xn]
(2) edge [out=200,in=160,looseness=7]   (2)
(1) edge[  left=15] (2)
(3) edge[  left=15] (2)
;
\end{tikzpicture}%
}}%
\hspace*{35mm}
\subfloat[Action graph.]{
\makebox[.2\textwidth]{%
\begin{tikzpicture}[x=1.5cm,y=1.2cm]
\TikzDecSyntInstance
\draw[xn]
(2) edge [out=200,in=160,looseness=7]  node[left]  {$1|3$} (2)
(1) edge node [below] {3}(2)
(3) edge node[below] {1}(2)
;
\end{tikzpicture}%
}}%
\caption{Action graph of the Parity protocol.}
\label{fig:synParity}
\end{figure}
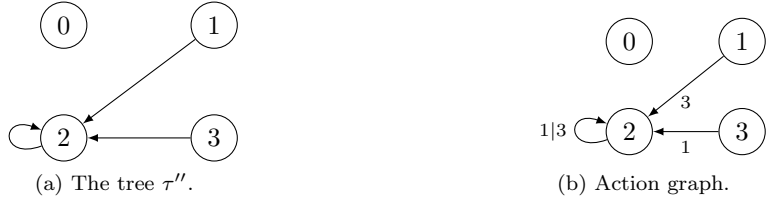

\noindent{\it Steps 9, 10, 11}:\ The remaining steps of the algorithm labels $\tau''$ (Figure \ref{fig:synParity}-(a)), which would result in the action graph in Figure \ref{fig:synParity}-(b). The synthesized  parameterized actions of the parity protocol are as follows:

\ \\
\hspace*{10mm}
\makebox[.8\textwidth]{%
$\begin{alignedat}{4}%
  &x_{i-1}{=}1 &{}\land{}& x_i{=}3 &{}\hspace*{1cm}\acteq{}\hspace*{1cm}& x_i \gets 2;
\\&x_{i-1}{=}3 &{}\land{}& x_i{=}1 &{}\hspace*{1cm}\acteq{}\hspace*{1cm}& x_i \gets 2;
\\&x_{i-1}{=}2 &{}\land{}&(x_i{=}1 \lor{} x_i{=}3) &{}\hspace*{1cm}\acteq{}\hspace*{1cm}& x_i \gets 2;
\end{alignedat}$
}

\subsection{Agreement Protocol}
\label{sec:agree}

Agreement is a fundamental problem in distributed computing. This section demonstrates how Algorithm  \ref{alg:SynLeadsTo} synthesizes a parameterized protocol on symmetric uni-rings that ensures agreement from a set of initial states. Specifically, we synthesize a protocol that meets $R \leadsto Q$, where $Q=\forall i \in \IntMod{N}: x_{i-1} = x_i$, and $R = \forall  i \in \IntMod{N}: (((x_i \ominus x_{i-1}) \mod 2= 0) \wedge (x_i \neq x_{i-1}))$.

\noindent{\it Steps 1 and 2}:\ Figure \ref{fig:gAgree} illustrates the locality graphs $G_R$ and $G_Q$, respectively for predicates $R$ and $Q$. We have four possible values for $\gamma$. Wlog, we let $\gamma$ be 1. 


\renewcommand{\TikzDecSyntInstance}{%
\tikzstyle{snod}=[draw,circle]
\tikzstyle{phantnod}=[inner sep=0pt]
\draw
( 0, 0) node[snod] (0) {$0$}
( 1, 0) node[snod] (1) {$1$}
( 0,-1) node[snod] (2) {$2$}
( 1,-1) node[snod] (3) {$3$}
;
\tikzstyle{lab}=[font=\scriptsize]
\tikzstyle{xn}=[draw,arrows=-latex,font=\scriptsize]}

\begin{figure}[h]
\centering
\subfloat[The locality graph $G_R$ for the state predicate $R$.]{
\makebox[.2\textwidth]{%
\begin{tikzpicture}[x=1.6cm,y=1.2cm]
\TikzDecSyntInstance
\draw[xn]
(1) edge[bend  left=15] (3)
(3) edge[bend  left=15] (1)
(0) edge[bend  left=15] (2)
(2) edge[bend  left=15] (0)
;
\end{tikzpicture}%
}}%
\hspace*{35mm}
\subfloat[The locality graph $G_Q$ for the state predicate $Q$.]{
\makebox[.2\textwidth]{%
\begin{tikzpicture}[x=1.5cm,y=1.2cm]
\TikzDecSyntInstance
\draw[xn]
(0) edge[out=200,in=160,looseness=7]  (0)
(1) edge[out=200,in=160,looseness=7]  (1)
(2) edge[out=200,in=160,looseness=7]  (2)
(3) edge[out=200,in=160,looseness=7]  (3)
;
\end{tikzpicture}%
}}%
\caption{Locality graphs of the Agreement protocol.}
\label{fig:gAgree}
\end{figure}
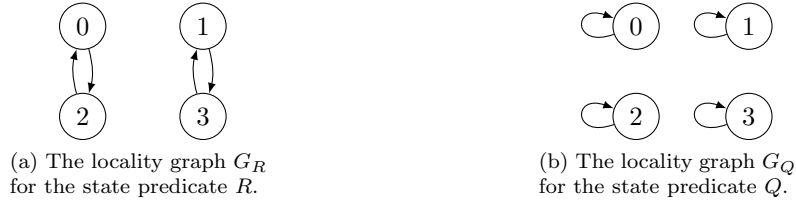

\noindent{\it Steps 3 and 4}:\ Since $G_Q$ has only self-loops on its vertices, $G'_Q$ would include just Vertex 1. Correspondingly, the spanning tree $\tau$ includes just a single vertex (i.e., 1).

\noindent{\it Steps 5, 6}:\ Since $V'_R$ is empty, $V'_{Rleaf}$ becomes empty too.

\noindent{\it Steps 7 and 8}:\ We have $V_Q - (V'_Q \cup V'_{Rleaf}) = \{0, 2, 3\}$. As a result, $\tau'$ would include arcs $(0,1)$ and $(2,1)$, but excludes $(3,1)$ because $r(3,1)$ holds (see Figure \ref{fig:synAgree}-(a)).  Executing Step 8 would result in including arcs $(3,2)$ and $(1,1)$, resulting in tree $\tau''$ (Figure \ref{fig:synAgree}-(b)).

\renewcommand{\TikzDecSyntInstance}{%
\tikzstyle{snod}=[draw,circle]
\tikzstyle{phantnod}=[inner sep=0pt]
\draw
( 0, 0) node[snod] (0) {$0$}
( 1, 0) node[snod] (1) {$1$}
( 0,-1) node[snod] (2) {$2$}
( 1,-1) node[snod] (3) {$3$}
;
\tikzstyle{lab}=[font=\scriptsize]
\tikzstyle{xn}=[draw,arrows=-latex,font=\scriptsize]}

\begin{figure}[h]
\centering
\subfloat[The tree $\tau'$.]{
\makebox[.2\textwidth]{%
\begin{tikzpicture}[x=2cm,y=1.5cm]
\TikzDecSyntInstance
\tikzstyle{pruned}=[dashed]
\draw[xn]
(0) edge[  left=15] (1)
(2) edge[  left=15] (1)
;
\end{tikzpicture}%
}}%
\hspace*{35mm}
\subfloat[The tree $\tau''$.]{
\makebox[.2\textwidth]{%
\begin{tikzpicture}[x=1.5cm,y=1.2cm]
\TikzDecSyntInstance
\draw[xn]
(1) edge [out=20,in=-20,looseness=7]   (1)
(0) edge[  left=15] (1)
(2) edge[  left=15] (1)
(3) edge[  left=15] (2)
;
\end{tikzpicture}%
}}%
\caption{Trees computed during synthesis of the Agreement  protocol.}
\label{fig:synAgree}
\end{figure}
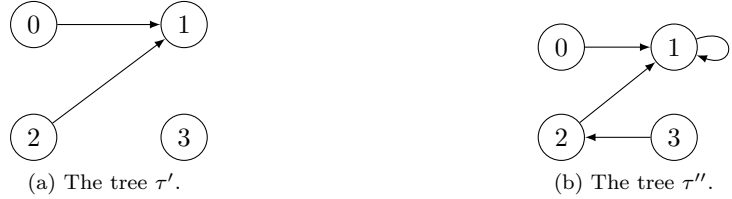

\noindent{\it Steps 9, 10, 11}:\ The remaining steps of the algorithm labels $\tau''$ (Figure \ref{fig:synAgree}-(b)), which would result in the action graph in Figure \ref{fig:actgAgree}.

\renewcommand{\TikzDecSyntInstance}{%
\tikzstyle{snod}=[draw,circle]
\tikzstyle{phantnod}=[inner sep=0pt]
\draw
( 0, 0) node[snod] (0) {$0$}
( 1, 0) node[snod] (1) {$1$}
( 0,-1) node[snod] (2) {$2$}
( 1,-1) node[snod] (3) {$3$}
;
\tikzstyle{lab}=[font=\scriptsize]
\tikzstyle{xn}=[draw,arrows=-latex,font=\scriptsize]}

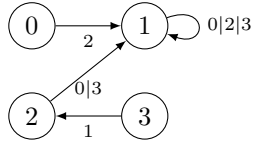
\begin{figure}[h]
\centering
\subfloat{
\makebox[.2\textwidth]{%
\begin{tikzpicture}[x=1.5cm,y=1.2cm]
\TikzDecSyntInstance
\draw[xn]
(1) edge [out=20,in=-20,looseness=7]  node[right]  {$0|2|3$} (1)
(0) edge node [below] {2}(1)
(2) edge node[below] {$0|3$}(1)
(3) edge node [below] {1}(2)
;
\end{tikzpicture}%
}}%
\caption{Action graph synthesized for Agreement .}
\label{fig:actgAgree}
\end{figure}

The synthesized  parameterized actions of the agreement protocol are as follows:

\makebox[.8\textwidth]{%
$\begin{alignedat}{4}%
  &x_{i-1}{=}0 &{}\land{}& x_i{=}2 &{}\acteq{}& x_i \gets 1;
\\&x_{i-1}{=}3 &{}\land{}& x_i{=}1 &{}\acteq{}& x_i \gets 2;
\\&x_{i-1}{=}2 &{}\land{}&(x_i{=}0 \lor{} x_i{=}3) &{}\acteq{}& x_i \gets 1;
\\&x_{i-1}{=}1 &{}\land{}&(x_i{=}0 \lor{} x_i{=}2\lor{} x_i{=}3) &{}\acteq{}& x_i \gets 1;
\end{alignedat}$
}

\section{Related Work}
\label{sec:related}

There is a diverse set of methods for the synthesis of fixed-size and parameterized programs/protocols. For example,  {\it program sketching} \cite{solar2006combinatorial,solar2013program} aims to automatically fill the holes in  an incomplete program.  {\it Example-based synthesis} \cite{so2017synthesizing} generates a program from a table of inputs and expected outputs. {\it Syntax-guided program synthesis}  \cite{alur2013syntax} constrains the search space of synthesis using a syntactic template of a program. 
 {\it Counterexample guided inductive synthesis} \cite{reynolds2015counterexample,abate2018counterexample} generates abstract programs and a verifier  provides   counterexamples for refining the synthesized abstract programs. Techniques for {\it automated completion} of distributed protocols \cite{udupa2013transit,alur2015automatic,udupa2016synthesis,alur2017automatic} mainly extend program sketching and synthesis-by-examples for distributed programs under strong/weak fairness. Existing automated techniques \cite{aae98,ftrtft2000,mcguire01,alithesis,icdcs07Borzoo,ftsyn,taasFarahat12,faghih2016specification,tpdsKlinkhamerE16,lmcsFaghihBTK18} for the addition of fault tolerance mainly enable the synthesis of fixed-size fault-tolerant protocols from their fault-intolerant versions.


Most existing methods for the verification of parameterized programs can be classified into  abstraction methods \cite{licsBhatCG95,fmsdIpD99,ijfcsEmersonN03}, SMT-based verification \cite{Ghilardi2010,conchon2012cubicle},   parameterized visual diagrams   \cite{sanchez2014parametrized},  network invariants \cite{avmfssWolperL89,kesten2002network,grinchtein2006inferring}, compositional model checking  \cite{mcmillan2001parameterized}, logic program transformations  and inductive verification methods 
\cite{roychoudhury2000verification,roychoudhury2001automated,roychoudhury2004inductively,fioravanti2013generalization}, regular model checking \cite{cavBouajjani00,etcsTouili01,concurAbdulla04}, proof spaces  \cite{farzan2016proving} and topology-specific verification \cite{matthews2016verifiable}.


 There are a variety of methods for  the synthesis of parameterized programs. For example, Attie and Emerson  \cite{attie1998synthesis} compose a pair of  representative processes (under weak fairness) to  reason about the global safety and local leads-to properties of a symmetric parameterized system. 
  Some researchers   \cite{dolev2013synchronous,bloem2016synthesis}  present methods for generating  parameterized  protocols for specific problems (e.g., counting) on specific topologies (e.g.,  clique \cite{dolev2013synchronous}).
Verification and synthesis methods based on {\it threshold automata}  \cite{lazic2018synthesis} take a sketch automaton  (whose transitions have incomplete guard conditions capturing the number of received messages), and complete the guards towards satisfying program specifications.



The closest work to this paper includes our previous work  on the  synthesis of self-stabilizing parameterized uni-rings \cite{tseEbnenasir19}, where a protocol is expected to recover to a set of states from {\em any} state. In the synthesis of self-stabilization, $R = true$ and $Q$ captures a set of legitimate states to which recovery is required. As such, we construct a spanning tree of the locality graph of $Q$ rooted at some $\gamma$. Such a spanning tree should include all values in the domain of $x_i$. By having just a self-loop on $\gamma$, we ensure that starting from any state in $\neg Q$, no livelocks will be reached (due to Theorem \ref{thm:LivelockIFF}). However, when $R \subset true$, it is not trivial how the synthesis should be conducted so global computations of a parameterized ring guarantee reachability of $Q$ only from global states in $R$. To address this challenge, we would need to identify local characterization of states that are reachable from $R$ as well as ensuring livelock-freedom only in states reachable from $R$. Algorithm \ref{alg:SynLeadsTo} succeeds in tackling these challenges. It is noteworthy to mention that a self-stabilizing solution may not be the best solution for $R \leadsto Q$ in cases where multiple leadsto properties must be satisfied.  For example, consider the case of two leadsto properties $R_1 \leadsto Q_1$ and $R_2 \leadsto Q_2$, where $R_1 \cap R_2 = \emptyset$. If one synthesizes a self-stabilizing protocol $p$ that recovers to $Q_1$, then $p$ certainly satisfies $R_1 \leadsto Q_1$ too. However, all hopes for revising $p$ towards satisfying $R_2 \leadsto Q_2$  are lost because $p$ instead  satisfies $R_2 \leadsto Q_1$. Using Algorithm \ref{alg:SynLeadsTo}, we ensure that $R_1 \leadsto Q_1$ is satisfied only from $R_1$, and not from states outside $R_1$. We are currently investigating conditions under which synthesis of  parameterized uni-rings that satisfy two leadsto properties $R_1 \leadsto Q_1$  and $R_2 \leadsto Q_2$ becomes possible.

\section{Conclusions and Future Work}
\label{sec:concl}

We investigated the problems of verifying and synthesizing parameterized protocols on symmetric unidirectional rings for leadsto properties $R \leadsto Q$. We showed that the verification problem remains undecidable even for constant-space and deterministic processes and for global state predicates $R$ and $Q$ that are formed by the conjunction of symmetric local state predicates; i.e., conjunctive predicates. We then presented  a somewhat surprising result that, synthesizing protocols that satisfy $R \leadsto Q$ on uni-rings is actually decidable! This is a significant result as both ring and leadsto are important aspects of distributed protocols. 
We are currently working on the implementation of our synthesis algorithm and conducting more interesting case studies such as cache coherence,  leader election, etc. Moreover, we plan to study the synthesis of leadsto on bidirectional rings, and other elementary topologies (e.g., mesh, torus).

\ \\

\noindent{\bf Acknowledgment}\\
The author would like to thank Aly Farahat for fruitful discussions.

\bibliographystyle{abbrv}
\bibliography{main}

\begin{thebibliography}{10}

\bibitem{abate2018counterexample}
A.~Abate, C.~David, P.~Kesseli, D.~Kroening, and E.~Polgreen.
\newblock Counterexample guided inductive synthesis modulo theories.
\newblock In {\em International Conference on Computer Aided Verification},
  pages 270--288. Springer, 2018.

\bibitem{concurAbdulla04}
P.~A. Abdulla, B.~Jonsson, M.~Nilsson, and M.~Saksena.
\newblock A survey of regular model checking.
\newblock In {\em CONCUR}, pages 35--48, 2004.

\bibitem{alur2013syntax}
R.~Alur, R.~Bodik, G.~Juniwal, M.~M. Martin, M.~Raghothaman, S.~A. Seshia,
  R.~Singh, A.~Solar-Lezama, E.~Torlak, and A.~Udupa.
\newblock Syntax-guided synthesis.
\newblock In {\em Formal Methods in Computer-Aided Design (FMCAD), 2013}, pages
  1--17. IEEE, 2013.

\bibitem{alur2015automatic}
R.~Alur, M.~Raghothaman, C.~Stergiou, S.~Tripakis, and A.~Udupa.
\newblock Automatic completion of distributed protocols with symmetry.
\newblock In {\em International Conference on Computer Aided Verification},
  pages 395--412. Springer, 2015.

\bibitem{alur2017automatic}
R.~Alur and S.~Tripakis.
\newblock Automatic synthesis of distributed protocols.
\newblock {\em ACM SIGACT News}, 48(1):55--90, 2017.

\bibitem{aae98}
P.~C. Attie, anish Arora, and E.~A. Emerson.
\newblock Synthesis of fault-tolerant concurrent programs.
\newblock {\em {ACM Transactions on Programming Languages and Systems
  (TOPLAS)}}, 26(1):125--185, 2004.

\bibitem{attie1998synthesis}
P.~C. Attie and E.~A. Emerson.
\newblock Synthesis of concurrent systems with many similar processes.
\newblock {\em ACM Transactions on Programming Languages and Systems (TOPLAS)},
  20(1):51--115, 1998.

\bibitem{ipdpsBernard09}
S.~Bernard, S.~Devismes, M.~G. Potop{-}Butucaru, and S.~Tixeuil.
\newblock Optimal deterministic self-stabilizing vertex coloring in
  unidirectional anonymous networks.
\newblock In {\em 23rd {IEEE} International Symposium on Parallel and
  Distributed Processing, {IPDPS} 2009, Rome, Italy, May 23-29, 2009}, pages
  1--8. {IEEE}, 2009.

\bibitem{licsBhatCG95}
G.~Bhat, R.~Cleaveland, and O.~Grumberg.
\newblock Efficient on-the-fly model checking for ctl*.
\newblock In {\em IEEE Symposium on Logic in Computer Science ({LICS})}, pages
  388--397, 1995.

\bibitem{bloem2016synthesis}
R.~Bloem, N.~Braud-Santoni, and S.~Jacobs.
\newblock Synthesis of self-stabilising and byzantine-resilient distributed
  systems.
\newblock In {\em International Conference on Computer Aided Verification},
  pages 157--176. Springer, 2016.

\bibitem{icdcs07Borzoo}
B.~Bonakdarpour and S.~S. Kulkarni.
\newblock Exploiting symbolic techniques in automated synthesis of distributed
  programs with large state space.
\newblock In {\em Proceedings of the 27th International Conference on
  Distributed Computing Systems}, pages 3--10, Washington, DC, USA, June 2007.
  IEEE Computer Society.

\bibitem{cavBouajjani00}
A.~Bouajjani, B.~Jonsson, M.~Nilsson, and T.~Touili.
\newblock Regular model checking.
\newblock In {\em CAV}, pages 403--418, 2000.

\bibitem{conchon2012cubicle}
S.~Conchon, A.~Goel, S.~Krstic, A.~Mebsout, and F.~Za{\i}di.
\newblock Cubicle: A parallel smt-based model checker for parameterized
  systems.
\newblock In {\em CAV}, pages 718--724. Springer, 2012.

\bibitem{de2011satisfiability}
L.~De~Moura and N.~Bj{\o}rner.
\newblock Satisfiability modulo theories: introduction and applications.
\newblock {\em Communications of the ACM}, 54(9):69--77, 2011.

\bibitem{dij}
E.~W. Dijkstra.
\newblock Self-stabilizing systems in spite of distributed control.
\newblock {\em Communications of the ACM}, 17(11):643--644, 1974.

\bibitem{dolev2013synchronous}
D.~Dolev, J.~H. Korhonen, C.~Lenzen, J.~Rybicki, and J.~Suomela.
\newblock Synchronous counting and computational algorithm design.
\newblock In {\em Symposium on Self-Stabilizing Systems}, pages 237--250.
  Springer, 2013.

\bibitem{alithesis}
A.~Ebnenasir.
\newblock {\em Automatic Synthesis of Fault-Tolerance}.
\newblock PhD thesis, Michigan State University, May 2005.

\bibitem{tseEbnenasir19}
A.~Ebnenasir and A.~Klinkhamer.
\newblock Topology-specific synthesis of self-stabilizing parameterized systems
  with constant-space processes.
\newblock {\em IEEE Transactions on Software Engineering}, 2019.
\newblock In Press.

\bibitem{ftsyn}
A.~Ebnenasir, S.~S. Kulkarni, and A.~Arora.
\newblock {FTSyn: A} framework for automatic synthesis of fault-tolerance.
\newblock {\em International Journal on Software Tools for Technology
  Transfer}, 10(5):455--471, 2008.

\bibitem{handb90}
E.~Emerson.
\newblock {\em Handbook of Theoretical Computer Science: Chapter 16, Temporal
  and Modal Logic}.
\newblock Elsevier Science Publishers B. V., 1990.

\bibitem{ijfcsEmersonN03}
E.~A. Emerson and K.~S. Namjoshi.
\newblock On reasoning about rings.
\newblock {\em International Journal of Foundations of Computer Science},
  14(4):527--550, 2003.

\bibitem{faghih2016specification}
F.~Faghih, B.~Bonakdarpour, S.~Tixeuil, and S.~Kulkarni.
\newblock Specification-based synthesis of distributed self-stabilizing
  protocols.
\newblock In {\em International Conference on Formal Techniques for Distributed
  Objects, Components, and Systems}, pages 124--141. Springer, 2016.

\bibitem{lmcsFaghihBTK18}
F.~Faghih, B.~Bonakdarpour, S.~Tixeuil, and S.~S. Kulkarni.
\newblock Automated synthesis of distributed self-stabilizing protocols.
\newblock {\em Logical Methods in Computer Science}, 14(1), 2018.

\bibitem{taasFarahat12}
A.~Farahat and A.~Ebnenasir.
\newblock A lightweight method for automated design of convergence in network
  protocols.
\newblock {\em {ACM Transactions on Autonomous and Adaptive Systems (TAAS)}},
  7(4):38:1--38:36, Dec. 2012.

\bibitem{icdcsFarahatE12}
A.~Farahat and A.~Ebnenasir.
\newblock Local reasoning for global convergence of parameterized rings.
\newblock In {\em {IEEE International Conference on Distributed Computing
  Systems (ICDCS)}}, pages 496--505, 2012.

\bibitem{farzan2016proving}
A.~Farzan, Z.~Kincaid, and A.~Podelski.
\newblock Proving liveness of parameterized programs.
\newblock In {\em Proceedings of the 31st Annual ACM/IEEE Symposium on Logic in
  Computer Science}, pages 185--196. ACM, 2016.

\bibitem{finkbeiner2013bounded}
B.~Finkbeiner and S.~Schewe.
\newblock Bounded synthesis.
\newblock {\em International Journal on Software Tools for Technology
  Transfer}, 15(5-6):519--539, 2013.

\bibitem{fioravanti2013generalization}
F.~Fioravanti, A.~Pettorossi, M.~Proietti, and V.~Senni.
\newblock Generalization strategies for the verification of infinite state
  systems.
\newblock {\em TPLP}, 13(2):175--199, 2013.

\bibitem{gascon2014synthesis}
A.~Gasc{\'o}n and A.~Tiwari.
\newblock Synthesis of a simple self-stabilizing system.
\newblock {\em arXiv preprint arXiv:1407.5392}, 2014.

\bibitem{Ghilardi2010}
S.~Ghilardi and S.~Ranise.
\newblock {\em MCMT: A Model Checker Modulo Theories}, pages 22--29.
\newblock Springer Berlin Heidelberg, Berlin, Heidelberg, 2010.

\bibitem{wssGouda01}
M.~Gouda.
\newblock The theory of weak stabilization.
\newblock In {\em 5th International Workshop on Self-Stabilizing Systems},
  volume 2194 of {\em Lecture Notes in Computer Science}, pages 114--123, 2001.

\bibitem{mcguire01}
M.~Gouda and T.~McGuire.
\newblock Correctness preserving transformations for network protocol
  compilers.
\newblock {\em Prepared for the Workshop on New Visions for Software Design and
  Productivity: Research and Applications}, 2001.

\bibitem{jpdcGouda96}
M.~G. Gouda and F.~F. Haddix.
\newblock The stabilizing token ring in three bits.
\newblock {\em Journal of Parallel and Distributed Computing}, 35(1):43--48,
  May 1996.

\bibitem{grinchtein2006inferring}
O.~Grinchtein, M.~Leucker, and N.~Piterman.
\newblock Inferring network invariants automatically.
\newblock In {\em Automated Reasoning}, pages 483--497. Springer, 2006.

\bibitem{spin97}
G.~Holzmann.
\newblock The model checker {SPIN}.
\newblock {\em IEEE Transactions on Software Engineering}, 23(5):279--295,
  1997.

\bibitem{fmsdIpD99}
C.~N. Ip and D.~L. Dill.
\newblock Verifying systems with replicated components in murphi.
\newblock {\em Formal Methods in System Design}, 14(3):273--310, 1999.

\bibitem{jacobs2012parameterized}
S.~Jacobs and R.~Bloem.
\newblock Parameterized synthesis.
\newblock In {\em International Conference on Tools and Algorithms for the
  Construction and Analysis of Systems}, pages 362--376. Springer, 2012.

\bibitem{kesten2002network}
Y.~Kesten, A.~Pnueli, E.~Shahar, and L.~Zuck.
\newblock Network invariants in action.
\newblock In {\em Concurrency Theory}, pages 101--115. Springer, 2002.

\bibitem{sssKlink13}
A.~Klinkhamer and A.~Ebnenasir.
\newblock Verifying livelock freedom on parameterized rings and chains.
\newblock In {\em International Symposium on Stabilization, Safety, and
  Security of Distributed Systems}, pages 163--177, 2013.

\bibitem{tpdsKlinkhamerE16}
A.~Klinkhamer and A.~Ebnenasir.
\newblock Shadow/puppet synthesis: {A} stepwise method for the design of
  self-stabilization.
\newblock {\em {IEEE} Trans. Parallel Distrib. Syst.}, 27(11):3338--3350, 2016.

\bibitem{livelock2019}
A.~Klinkhamer and A.~Ebnenasir.
\newblock On the verification of livelock-freedom and self-stabilization on
  parameterized rings.
\newblock {\em {ACM Transactions on Computational Logic}}, 2019.
\newblock Accepted.

\bibitem{fsenKlink17}
A.~P. Klinkhamer and A.~Ebnenasir.
\newblock Synthesizing parameterized self-stabilizing rings with constant-space
  processes.
\newblock In {\em International Conference on Fundamentals of Software
  Engineering}, pages 100--115. Springer, 2017.

\bibitem{ftrtft2000}
S.~S. Kulkarni and A.~Arora.
\newblock Automating the addition of fault-tolerance.
\newblock In {\em Formal Techniques in Real-Time and Fault-Tolerant Systems},
  pages 82--93, London, UK, 2000. Springer-Verlag.

\bibitem{lazic2018synthesis}
M.~Lazic, I.~Konnov, J.~Widder, and R.~Bloem.
\newblock Synthesis of distributed algorithms with parameterized threshold
  guards.
\newblock In {\em LIPIcs-Leibniz International Proceedings in Informatics},
  volume~95. Schloss Dagstuhl-Leibniz-Zentrum fuer Informatik, 2018.

\bibitem{matthews2016verifiable}
O.~Matthews, J.~Bingham, and D.~J. Sorin.
\newblock Verifiable hierarchical protocols with network invariants on
  parametric systems.
\newblock In {\em Formal Methods in Computer-Aided Design (FMCAD), 2016}, pages
  101--108. IEEE, 2016.

\bibitem{mcmillan2001parameterized}
K.~L. McMillan.
\newblock Parameterized verification of the flash cache coherence protocol by
  compositional model checking.
\newblock In {\em Correct hardware design and verification methods}, pages
  179--195. Springer, 2001.

\bibitem{pr90}
A.~Pnueli and R.~Rosner.
\newblock Distributed reactive systems are hard to synthesis.
\newblock In {\em Proceedings of 31st IEEE Symposium on Foundation of Computer
  Science}, pages 746--757, Washington, DC, USA, 1990. IEEE Computer Society.

\bibitem{reynolds2015counterexample}
A.~Reynolds, M.~Deters, V.~Kuncak, C.~Tinelli, and C.~Barrett.
\newblock Counterexample-guided quantifier instantiation for synthesis in smt.
\newblock In {\em International Conference on Computer Aided Verification},
  pages 198--216. Springer, 2015.

\bibitem{roychoudhury2000verification}
A.~Roychoudhury, K.~N. Kumar, C.~Ramakrishnan, I.~Ramakrishnan, and S.~A.
  Smolka.
\newblock Verification of parameterized systems using logic program
  transformations.
\newblock In {\em Tools and Algorithms for the Construction and Analysis of
  Systems}, pages 172--187. Springer, 2000.

\bibitem{roychoudhury2001automated}
A.~Roychoudhury and I.~Ramakrishnan.
\newblock Automated inductive verification of parameterized protocols?
\newblock In {\em Computer Aided Verification}, pages 25--37. Springer, 2001.

\bibitem{roychoudhury2004inductively}
A.~Roychoudhury and I.~Ramakrishnan.
\newblock Inductively verifying invariant properties of parameterized systems.
\newblock {\em Automated Software Engineering}, 11(2):101--139, 2004.

\bibitem{sanchez2014parametrized}
A.~S{\'a}nchez and C.~S{\'a}nchez.
\newblock Parametrized verification diagrams.
\newblock In {\em Temporal Representation and Reasoning (TIME), 2014 21st
  International Symposium on}, pages 132--141. IEEE, 2014.

\bibitem{so2017synthesizing}
S.~So and H.~Oh.
\newblock Synthesizing imperative programs from examples guided by static
  analysis.
\newblock In {\em International Static Analysis Symposium}, pages 364--381.
  Springer, 2017.

\bibitem{solar2013program}
A.~Solar-Lezama.
\newblock Program sketching.
\newblock {\em International Journal on Software Tools for Technology
  Transfer}, 15(5-6):475--495, 2013.

\bibitem{solar2006combinatorial}
A.~Solar-Lezama, L.~Tancau, R.~Bodik, S.~Seshia, and V.~Saraswat.
\newblock Combinatorial sketching for finite programs.
\newblock {\em ACM Sigplan Notices}, 41(11):404--415, 2006.

\bibitem{suzuki1988proving}
I.~Suzuki.
\newblock Proving properties of a ring of finite-state machines.
\newblock {\em Information Processing Letters}, 28(4):213--214, July 1988.

\bibitem{etcsTouili01}
T.~Touili.
\newblock Regular model checking using widening techniques.
\newblock {\em Electronic Notes in Theoretical Computer Science},
  50(4):342--356, 2001.

\bibitem{udupa2016synthesis}
A.~Udupa.
\newblock {\em Synthesis of distributed protocols from scenarios and
  specifications}.
\newblock PhD thesis, University of Pennsylvania, 2016.

\bibitem{udupa2013transit}
A.~Udupa, A.~Raghavan, J.~V. Deshmukh, S.~Mador-Haim, M.~M. Martin, and
  R.~Alur.
\newblock Transit: specifying protocols with concolic snippets.
\newblock In {\em ACM SIGPLAN Notices}, volume~48, pages 287--296. ACM, 2013.

\bibitem{varPhD93}
G.~Varghese.
\newblock {\em Self-stabilization by local checking and correction}.
\newblock PhD thesis, MIT, 1993.

\bibitem{avmfssWolperL89}
P.~Wolper and V.~Lovinfosse.
\newblock Verifying properties of large sets of processes with network
  invariants.
\newblock In {\em International Workshop on Automatic Verification Methods for
  Finite State Systems}, pages 68--80, 1989.

\end{thebibliography}

\end{document}